\documentclass[pra,reprint, showpacs, a4paper,nofootinbib]{revtex4-2}
\usepackage{graphicx}
\usepackage{amsmath}
\usepackage{siunitx}
\usepackage{amssymb}
\usepackage{color}
\usepackage[latin1]{inputenc}
\usepackage{mathrsfs}
\usepackage{mathtools}
\usepackage{esint}
\usepackage{multirow}

\renewcommand{\vec}[1]{\boldsymbol{#1}}

\usepackage{wasysym}

\newcommand{\om}[1]{\omega_{\text{#1}}}
\newcommand{\vphi}[1]{\varphi_{\text{#1}}}
\newcommand{\ta}[1]{\tau_{\text{#1}}}

\newcommand{\define}{\coloneqq}
\definecolor{orange}{rgb}{1,0.647,0}

\usepackage{tikz}
\newcommand{\pgftextcircled}[1]{
	\setbox0=\hbox{#1}%
	\dimen0\wd0%
	\divide\dimen0 by 2%
	\begin{tikzpicture}[baseline=(a.base)]%
	\useasboundingbox (-\the\dimen0,0pt) rectangle (\the\dimen0,1pt);
	\node[circle,draw,outer sep=0pt,inner sep=0.1ex] (a) {#1};
	\end{tikzpicture}
}

\begin{document}

\title{Determination of atomic multiphoton ionization phases 
	 \\by trichromatic multichannel wave packet interferometry}

\date{\today}
\author{K. Eickhoff, D. Köhnke, L. Feld, T. Bayer, M. Wollenhaupt}
\affiliation{Carl von Ossietzky Universit\"at Oldenburg, Institut f\"ur Physik, Carl-von-Ossietzky-Stra\ss e 9-11, D-26129 Oldenburg Germany}

\begin{abstract}
We present a multichannel photoelectron interferometry technique based on trichromatic pulse shaping for unambiguous determination of quantum phases in multiphoton ionization (MPI) of potassium atoms. The colors of the laser field are chosen to produce three energetically separated photoelectron interferograms in the continuum. While the red pulse is two-photon resonant with the $3d$-state resulting in a (2+1) resonance-enhanced MPI (REMPI), a (1+2) REMPI occurs via the non-resonant intermediate $4p$-state with an initial green or blue pulse. We show that ionization via a non-resonant intermediate state lifts the degeneracy of photoelectron interferograms from pathways consisting of permutations of the colors. The analysis of the interferograms reveals a phase shift of $\pm \pi/2$ depending on the sign of the detunings in the (1+2) REMPI pathways. In addition, we demonstrate that the photoionization time delay in the resonant (2+1) REMPI pathway gives rise to a linear spectral phase in the photoelectron spectra. Insights into the underlying MPI processes are gained through an analytic perturbative description and numerical simulations of a trichromatic driven three level system coupled to the continuum.\\
\end{abstract}


\maketitle

\section{Introduction}
Being at the heart of atomic, molecular, and optical physics, photoionization has been studied for over a century, both experimentally and theoretically. Initially, Albert Einstein gave a theoretical interpretation of the photoelectric effect in his seminal 1905 publication \cite{Einstein:1905:AP:165}, where he envisioned the possibility of processes in which an ``energy quantum of emitted light can obtain its energy from multiple incident energy quanta''. Subsequently, a quantum mechanical description of two-photon transitions was given by Maria Göppert-Mayer in 1931 \cite{Goeppert-Mayer:1931:AP:273} paving the way to our modern understanding of multiphoton ionization (MPI). Sophisticated MPI schemes combined with highly differential detection techniques for the time-, energy- and angle-resolved measurements have established the field of nonlinear photoelectron spectroscopy yielding numerous applications in fundamental and applied physics, as described in monographs \cite{Lambropoulos:1976:87,Faisal:1987,Fujimura:2003:199,Letokhov:1987,Kitzler:2016,Lin:2016,Bauer:2017,Hockett:2018a,Yamanouchi:2021}
and review articles \cite{Fedorov:1989:PQE:1,Mainfray:1991:RPP:1333,Doerner:2000:PR:95,Wollenhaupt:2005:ARPC:25,Reid:2012:MP:131}. Attosecond electron dynamics
\cite{Goulielmakis:2010:Nature:739,Villeneuve:2017:Science:1150}, coherent control of atomic and molecular dynamics by shaped laser pulses  \cite{Rice:2000:456,Goswami:2003:PR:385,Brif:2010:NJP:075008,Shapiro:2012:562,Wollenhaupt:2015:63,Trallero-Herrero:2006:PRL:063603} and the generation of free electron vortices \cite{NgokoDjiokap:2015:PRL:113004,Yuan:2016:PRA:053425,Pengel:2017:PRL:053003,Kerbstadt:2019:NC:658,Eickhoff:2021:FP:444} are examples for  the wide variety of modern topics in atomic and molecular MPI. In addition to measuring the photoelectron probability density distribution the use of interferometric techniques \cite{Wollenhaupt:2002:PRL:173001} has made it possible to determine the phase of photoelectron wave packets. Quantum mechanical phases are key to coherent control \cite{Warren:1993:Science:1581,Rabitz:2000:Science:824,Rabitz:2011:FD:415,Weinacht:1999:Nature:233,Kaufman:2020:PRL:053202} and essential for the complete characterization of the wave function \cite{Leichtle:1998:PRL:1418,Averbukh:1999:PRA:2163,deMorissonFaria:2020:RPP:034401}. Knowledge about the phase of photoelectron wave functions from atomic \cite{Isinger:2017:Science:893,Swoboda:2010:PRL:103003,Drescher:2022:PRA:L011101} and molecular \cite{Haessler:2009:PRA:011404,Vos:2018:Science:1326,Grundmann:2020:Science:339} photoionization processes provides insights into the underlying dynamics \cite{Gruson:2016:Science:734} accompanied with photoemission phases and time delays \cite{Hofmann:2019:JMO:1052,Sokolovski:2018:CP:1,Hockett:2016:JPBAMOP:095602,Harth:2019:PRA:023410,Dahlstroem:2013:CP:53,Schultze:2010:Science:1685,Dahlstroem:2012:JPB:183001}. In those measurements, the superposition of multiple partial waves gave rise to phase-sensitive interference patterns that allowed the quantum mechanical phase to be reconstructed. In general, the exact shape of an interferogram is determined by the ionizing radiation's optical phase and the additional quantum phase due to the ionization dynamics. To distinguish between these two phases, in the following, the phase introduced by laser radiation will be referred to as \textit{optical phase}, while the quantum mechanical phase resulting from ionization dynamics will be referred to as \textit{MPI-phase}. Very recently, building on the results presented in \cite{Swoboda:2010:PRL:103003}, MPI-phases from multiple intermediate states were investigated in the photoionization of helium atoms \cite{Drescher:2022:PRA:L011101}. \\

In this paper, we demonstrate a trichromatic MPI scheme to decouple the MPI-phase measurement from the optical phase control for an unambiguous determination of the MPI-phase. We exemplify our scheme on the trichromatic three-photon ionization of potassium atoms to investigate the MPI-phases due to the intermediate resonances. It has been shown that interference patterns in photoelectron spectra from multichromatic MPI, such as bichromatic photoelectron vortices, are sensitive to the carrier-envelope phase (CEP) \cite{Kerbstadt:2019:NC:658}. However, we have recently demonstrated a pulse-shaper-based trichromatic MPI scheme for quantum state holography \cite{Eickhoff:2021:PRA:052805} in which the photoelectron hologram is sensitive to the MPI-phases and the relative optical phases between the colors in the trichromatic pulse sequence but insensitive to the CEP. In this work, we use temporally overlapping trichromatic fields with the central frequencies of the red, green and blue spectral band chosen such that $2\omega_\text{g} = \omega_\text{r}+\omega_\text{b}$. Due to this frequency condition, the laser pulses create free electron wave packets forming interferograms in three different photoelectron kinetic energy windows, referred to as interference channels. By design, the interferograms consist of wave packets from (1+2) resonance enhanced MPI (REMPI) via the $4p$-state or (2+1) REMPI via the $3d$-state. The observed interference patterns are sensitive to the MPI-phases as well as the relative optical phases. Motivated by the concerted generation of multiple interferograms in one photoelectron spectrum, the technique is termed \textit{trichromatic multichannel wave packet interferometry}. Due to the temporal overlap of the fields, the observed interferograms also map the phases introduced by transiently populated bound atomic states. The MPI-phases are extracted from energy-resolved photoelectron spectra measured as a function of the relative optical phase of the green spectral component. In the resulting phase maps, the quantum phases manifest as relative phase shifts between the interferograms in the different channels. The interpretation of the experimental results is supported by the analysis of a model system consisting of a three-level atom coupled to the continuum. To simulate laser-induced bound and ionization dynamics we study both, the analytic perturbative description and the numeric solution of the time-dependent Schrödinger equation (TDSE).

\section{Physical system and Theory}
\label{sec:system:theory}
In this section, we present the physical system studied by our shaper-based multichannel interferometry scheme. We provide a theoretical description of the measured photoelectron momentum distributions (PMDs) with emphasis on their dependence on the optical and MPI-phases. Based on the theoretical model, we analyze our experimental results in Sec.~\ref{sec:results_discussion}.\\

\subsection{Physical system}\label{sec:phys_sys}
\begin{figure*}[t]
	\includegraphics[width=\linewidth]{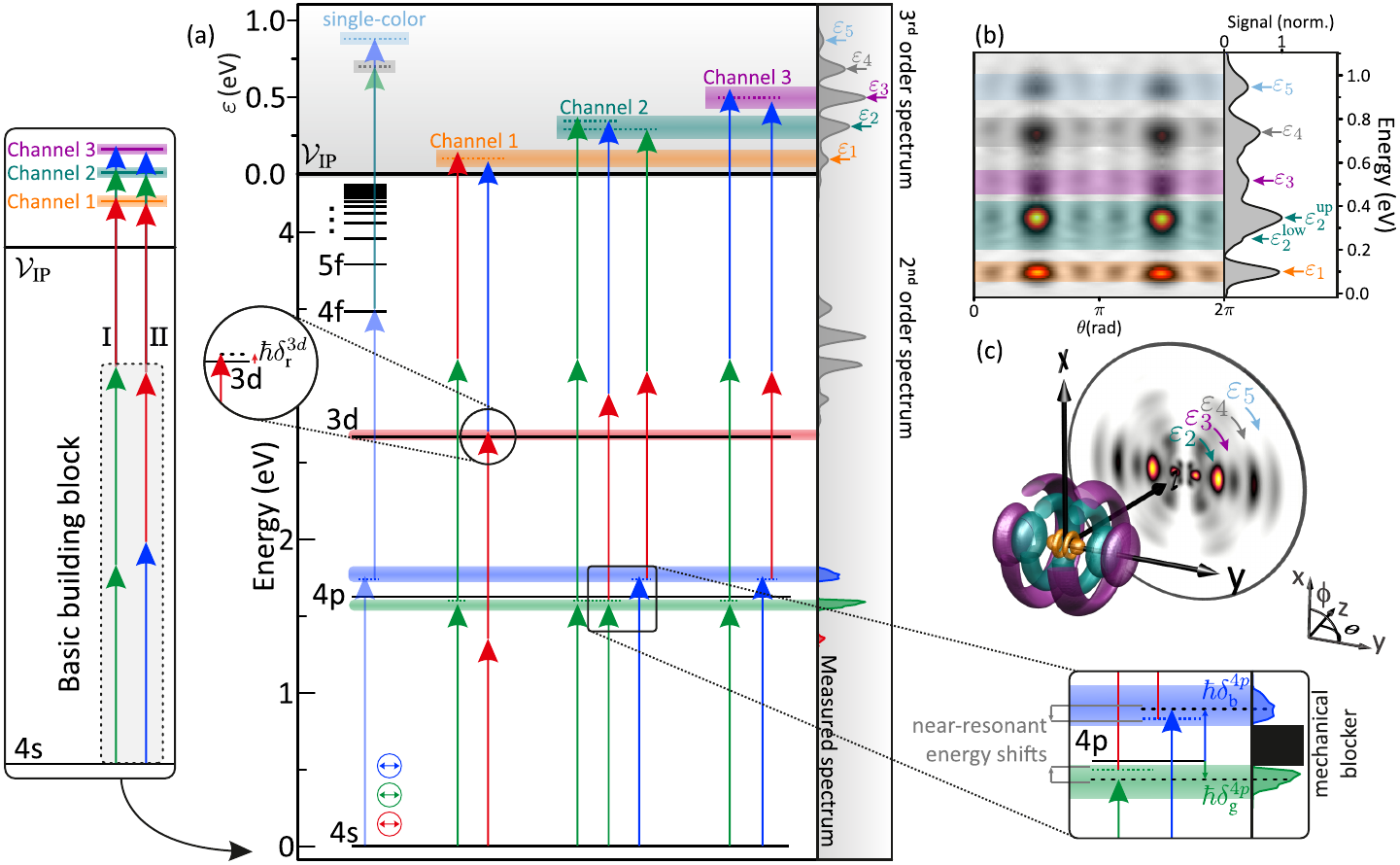}
	\caption{
		(a) Excitation scheme of potassium atoms interacting perturbatively with a linearly polarized trichromatic field for multichannel interferometry together with the experimental multiphoton spectrum. We use temporally overlapping trichromatic pulses with frequencies $2\omega_\text{g} = \omega_\text{r}+\omega_\text{b}$ (black dashed box in the left figure) to achieve interference of photoelectrons by MPI with different combinations of red, green and blue photons. The interference leads to three energetically separated photoelectron interferograms in the interference channels 1-3 (highlighted in orange, petrol and violet). The frequencies are spectrally confined to be either two-photon resonant with the $3d$-state, i.e. $2 \om{r} \simeq \om{3d}$, or one-photon near- but non-resonant with the $4p$-state, i.e., $\om{g} \lesssim \om{4p} < \om{b}$ (right bottom inset). 
		The near-resonant excitation of the $4p$-state lifts the degeneracy of the  frequency mixing pathways and causes the second channel to split into two bands ($\varepsilon^{\text{low}}_2$ and $\varepsilon^{\text{up}}_2$) around $\varepsilon_2$. (b) Measured and Abel-inverted photoelectron signals in polar coordinates. (c) Measured and tomographically reconstructed 3D PMD with color-coded interference channels at $\varepsilon_1$-$\varepsilon_3$. }
	\label{fig:ExcitationScheme}
\end{figure*}

The excitation scheme of potassium atoms interacting perturbatively with parallel linearly polarized (PLP) trichromatic fields is shown in Fig.~\ref{fig:ExcitationScheme}(a). In the experiment, we use temporally overlapping trichromatic pulses to produce the interference of multiple MPI pathways from \textit{different} combinations of red (r), green (g) and blue (b) photons, i.e., single-color and frequency mixing contributions \cite{Kerbstadt:2017:NJP:103017} in the photoelectron spectrum. Usually, in multichromatic MPI, only permutations of a given color combination connect the same initial and final states. However, by appropriate choice of the field's frequencies, even different color combinations give rise to interference in several energetically separated final states. In the experiment, the photon energies are chosen to ensure the same number of photons (\textit{three}) not only in each frequency mixing pathway but also in the green single-color pathway. Consequently, the photoelectron interferograms are insensitive to CEP variations, but are controlled by the relative optical phases. To maximize the generation of the frequency mixing signals, all three colors overlap in time. The interference of frequency mixing pathways consisting of different photon combinations is observed when the central frequencies $\omega_j$ and the time delays $\tau_j$ of the trichromatic field are chosen such that
\begin{align}
\label{eq:multichannelcondition1}
2  \omega_\mathrm{g}&=  \omega_\mathrm{r}+ \omega_\mathrm{b} < \omega_\text{IP},\\
\label{eq:multichannelcondition2}
\tau_\mathrm{r}&= \tau_\mathrm{g} = \tau_\mathrm{b},
\end{align}
where $\mathcal{V}_\text{IP} = \hbar \omega_\text{IP}$ describes the ionization potential. The spectral and temporal conditions \eqref{eq:multichannelcondition1} and \eqref{eq:multichannelcondition2} form the basic building block of the presented scheme which is visualized by the dashed box in the left inset of Fig.~\ref{fig:ExcitationScheme}(a). When both conditions are met, the trichromatic field simultaneously generates three energetically separated photoelectron interferograms in the continuum. In the following, the corresponding energy windows centered around $\varepsilon_1$, $\varepsilon_2$ and $\varepsilon_3$ will be referred to as \textit{interference channels} (cf. Fig.~\ref{fig:ExcitationScheme}(a)). To study the MPI-phases induced by intermediate resonant states, the frequencies are further chosen to be either two-photon resonant with the $3d$-state, i.e. $2 \om{r} = \om{3d}$, or one-photon near-resonant with the $4p$-state, i.e., $\om{g} \lesssim \om{4p} < \om{b}$. Since the (near-)resonant excitation is more efficient, the $4p$ and $3d$ intermediate states select the resonance-enhanced pathways, listed in Tab.~\ref{tab:NotationPaths}.\\

For example, the interferogram in channel 1 is created by the interference of photoelectrons via the resonance-enhanced MPI pathways $\mathcal{R}_\mathrm{I}$ and $\mathcal{R}_\mathrm{II}$. The pathway $\mathcal{R}_\mathrm{I}$, denoted as (ggr), describes the absorption of two green (gg) photons and one red (r) photon in this order. Similarly, the pathway $\mathcal{R}_\mathrm{II}$, denoted as (rrb), describes the absorption of two red (rr) photons and one blue (b) photon. 
The other resonance-enhanced pathways leading to the channels 2 and 3 are labeled in the same way. Even though the color combinations of the individual MPI pathways are quite different, the total optical phase difference $\Delta\varphi$ introduced by the respective laser pulses is the same for all interferograms. The perturbative description of the REMPI in Sec.~\ref{sec:pert_description} shows that this phase difference is $\Delta\varphi=2\vphi{g}-\vphi{r}-\vphi{b}$, where $\vphi{r}$, $\vphi{g}$ and $\vphi{b}$ are the relative optical phases of the red, green and blue pulse, respectively (cf. Eq.~\eqref{eq:phase_difference}). Therefore, the variation of $\Delta\varphi$ induces a uniform amplitude modulation of the interferograms in each channel. By scanning the relative optical phase $\vphi{g}$ and measuring energy-resolved photoelectron spectra
we map out the amplitude modulation. In the following, these phase-resolved measurements are denoted as \textit{phase maps}. Because the influence of the total optical phase difference $\Delta\varphi$ and the CEP is the same in each interferogram, relative phase shifts between the recorded modulations in the phase maps are exclusively attributed to differences in the MPI-phases acquired in the corresponding pathways. Thus, in this trichromatic scheme the recorded phase maps provide direct access to the MPI-phases via the observed phase shifts between the interferograms.\\
 
In the experiment we use temporally overlapping trichromatic PLP laser fields with central wavelengths of $\lambda_\text{r} = \SI{925}{nm}$ (red band), $\lambda_\text{g} = \SI{797}{nm}$ (green band) and $\lambda_\text{b} = \SI{708}{nm}$ (blue band). The corresponding frequencies $\omega_j$ meet the frequency condition in Eq.~\eqref{eq:multichannelcondition1}. To study the MPI-phase from resonant excitation, the red pulse is two-photon resonant with the $3d$-state \cite{Kramida:2018} with the very small detuning ${\hbar \delta_\mathrm{r}^{3d}=2 \hbar \omega_\mathrm{r}- \hbar \omega_{3d}\approx \SI{10}{meV}}$ (cf. left magnification in Fig.~\ref{fig:ExcitationScheme}(a)). The spectral full width at half maximum (FWHM) of the red pulse is $\hbar \Delta \omega_r \approx \SI{30}{meV}$. In contrast, the MPI-phases due to the transient population dynamics in the non-resonant $4p$-state are studied with the green ($\hbar \Delta \omega_\text{g} \approx \SI{60}{meV} $) and blue ($\hbar \Delta \omega_\text{b} \approx \SI{70}{meV}$) pulses. To eliminate the spectral intensity of the pulse at the transition frequency $\omega_{4p}$ \cite{Kramida:2018} a mechanical blocker is inserted into the Fourier plane of our pulse shaper (cf. bottom right inset of Fig.~\ref{fig:ExcitationScheme}(a) and Fig.~\ref{fig:Strategy}(a)). This procedure guarantees non-resonant excitation of the $4p$-state despite the small detuning. As a consequence, the spectrum of the green pulse is slightly asymmetric. The detunings of the green and blue pulse are ${\hbar \delta_\mathrm{g}^{4p}=\hbar \omega_\mathrm{g}- \hbar \omega_{4p} \approx \SI{-50}{meV}}$ and ${\hbar \delta_\mathrm{b}^{4p}=\hbar \omega_\mathrm{b}-\hbar \omega_{4p} \approx \SI{140}{meV}}$, respectively (cf. right magnification in Fig.~\ref{fig:ExcitationScheme}(a)). Thus, our trichromatic scheme allows us to study MPI-phases from excitation of two different intermediate states either via resonant or via red- ($\delta_\mathrm{g}^{4p} <0$) and blue-detuned ($\delta_\mathrm{b}^{4p} >0$) near-resonant pathways simultaneously.\\

In our theoretical model, the atom is described by a three-level system coupled to the continuum. The bound part consists of the ground state $\vert g \rangle$ and the two (near-)resonant intermediate states $\vert a \rangle$ and $\vert b \rangle$. The final state $\vert f \rangle$ with variable energy describes the free electron in the continuum (for details see Sec.~\ref{sec:pert_description}). Due to the choice of the photon energies described above, the photoelectron interferograms in channels 1 and 3 are centered around the energies $\varepsilon_1 = \SI{0.10}{eV}$ and $\varepsilon_3 = \SI{0.50}{eV}$. Generally, pathways composed of the same color combinations are energetically degenerate since they differ only by permutations. However, the transient population of the near-resonant $4p$-state introduces a spectral shift. In particular, since the green pulse is slightly red-detuned ($\delta_\mathrm{g}^{4p} <0$) with respect to the $4s \rightarrow 4p$ transition, its transient population follows the laser field envelope with an additional temporal phase. As a consequence, the photoelectron signals mapping the $4p$-state are slightly shifted upwards in energy. For the same reason, wave packets created via pathways starting with a blue photon ($\delta_\mathrm{b}^{4p} >0$) are slightly shifted downwards in energy. Thus, near- but non-resonant excitation of the $4p$-state lifts the degeneracy of the involved frequency mixing contributions introducing an energetic splitting between the pathways with an initial green photon and those starting with a blue photon. In the interference channels 1 and 3 only the upper signals are observed since the multiplicity of the green-green-red and green-green-blue pathways favors the ones with a leading green photon. In channel 2 the red-green-blue pathway around $\varepsilon_2 = \hbar (\om{r} + \om{g} + \om{b}) = \SI{0.31}{eV}$ exhibits both an up-shifted signal centered around $\varepsilon^{\text{up}}_2$ from the pathways with an initial green photon and a down-shifted signal centered around $\varepsilon^{\text{low}}_2$ from the pathways with an initial blue photon. Note that this energetic separation is not to be confused with the Stark shift or the Autler-Townes splitting \cite{Wollenhaupt:2003:PRA:015401}. It is intensity-independent and fully consistent with the perturbative description of the excitation (see Sec.~\ref{sec:pert_description} and Appendix \ref{app:Pert_Description} for more details). Indeed, the energetic separation permits us to investigate both, the red- and the blue-detuned near- but non-resonant excitation of the $4p$-state, independently in a single interference channel (cf. bottom right inset of Fig.~\ref{fig:ExcitationScheme}(a)). In addition to the above mentioned interferograms, we observe photoelectron signals at $\varepsilon_4 = \SI{0.73}{eV}$ and $\varepsilon_5 = \SI{0.93}{eV}$. While the contribution at $\varepsilon_4$ originates from a frequency mixing pathway with two blue and one green photon, the one at $\varepsilon_5$ arises from the blue single-color pathway, depicted in Fig.~\ref{fig:ExcitationScheme}(a). The former contribution is used for \textit{in-situ} temporal pulse characterization in Sec.~\ref{sec:exp_strategy}. Figure~\ref{fig:ExcitationScheme}(b) depicts a section through the $y$-$z$-plane in polar representation and the energy-resolved spectrum obtained by integration over the angular coordinate of the measured PMD, highlighting the different contributions at $\varepsilon_1$-$\varepsilon_5$. Figure~\ref{fig:ExcitationScheme}(c) shows the measured and Abel-inverted \cite{Garcia:2004:RSI:4989} three dimensional PMDs of the three color-coded energetically nested $f(m=0)$-type interferograms (details see Sec.~\ref{sec:experiment}). In this figure, the 3D PMDs of the contributions around $\varepsilon_4$ and $\varepsilon_5$ are omitted for clarity. 

\subsection{Perturbative description}\label{sec:pert_description}

In this section, we provide a theoretical description of the photoelectron wave packets created in the multichannel interferometry scheme. We focus on the spectral phases in the different interference channels introduced by the (near-)resonant excitation in the MPI process. First, we introduce the trichromatic laser field in the temporal and spectral representation. Using perturbation theory, we derive a general expression for the photoelectron density, based on a multilevel model to calculate the MPI-phases in the relevant pathways.\\ 

The trichromatic PLP field is fully characterized by the scalar electric fields of each color. The total electric field (negative frequency analytic signal \cite{Diels:1996}) is therefore given by
\begin{equation}\label{eq:temporal_field}
E^-(t) = \sum_{j\in \lbrace \text{r},\text{g},\text{b} \rbrace } \mathcal{E}_j(t-\tau_j)e^{-i(\omega_j t-\varphi_j)},
\end{equation}
with the temporal pulse envelopes $\mathcal{E}_j(t)$ shifted in time by the delay $\tau_j$, the central frequency $\omega_j$ and the relative optical phase $\varphi_j$. In Eq.~\eqref{eq:temporal_field} the CEP of the field components is omitted because in our scheme the phase in the photoelectron spectrum introduced by the CEP is identical for all MPI pathways. The corresponding spectrum reads
\begin{equation}
\tilde{E}^-(\omega) = \sum_{j\in \lbrace \text{r},\text{g},\text{b}  \rbrace } \tilde{\mathcal{E}}_j(\omega + \omega_j)e^{i( \varphi_i - (\omega + \omega_i)\tau_i)}
\end{equation}
with the spectra of the envelopes $\tilde{\mathcal{E}}_j(\omega)$. Again, the three central frequencies $\omega_j$ are chosen such that the energy of two green photons equals the energy of one red plus one blue photon as described by Eq.~\eqref{eq:multichannelcondition1}. Simultaneous absorption of another red, green or blue photon in both processes provides the photon energy to overcome the ionization potential and gives rise to the interference channels $l=1$, $2$ and $3$, respectively (cf. visualization of the basic building block in the left inset of Fig.~\ref{fig:ExcitationScheme}(a)). Therefore, each interference channel $l$ consists of two pathways \textbf{I} and \textbf{II} either with at least two green photons (\textbf{I}) or with at least one red and one blue photon (\textbf{II}). Both pathways contain a photon of the additional color $x_l$ with $\boldsymbol{x} =  [\text{r},\text{g},\text{b}] $, e.g., $x_1 = \text{r}$ for the additional red photon in channel 1. Due to the temporal overlap of the three colors (cf. Eq.~\eqref{eq:multichannelcondition2}), every permutation of the involved photons gives rise to another MPI pathway. We can write both pathways as a set containing all permutations of the involved colors, i.e., $\mathcal{Q}^{(l)}_\text{\textbf{I}} = \text{Sym}(\lbrace \text{g},\text{g},x_l \rbrace )$ and $\mathcal{Q}^{(l)}_\text{\textbf{II}} = \text{Sym}(\lbrace \text{r},\text{b},x_l \rbrace )$ where Sym denotes the symmetric group. In general the resulting photoelectron interferogram in channel $l$ is given by the superposition of all possible partial wave functions created via the pathways $\mathcal{Q}^{(l)}_\text{\textbf{I}}$ and $\mathcal{Q}^{(l)}_\text{\textbf{II}}$. However, as discussed in  Sec.~\ref{sec:phys_sys}, the intermediate resonances in the potassium atom select the relevant pathways. Therefore, the description focuses on the superposition of relevant resonance-enhanced quantum pathways $\mathrm{I}\in\mathcal{Q}_\mathrm{\textbf{I}}^{(l)}$ and $\mathrm{II}\in\mathcal{Q}_\mathrm{\textbf{II}}^{(l)}$, where the ordering of the colors is fixed due to the resonances. As an example, we consider the superposition of the (ggb) and (brb) pathways in channel 3 (cf. Tab.~\ref{tab:NotationPaths}). Hence, we assume each interferogram $\Psi^{(l)}$ to be composed of two partial wave functions $\Psi^{(l)} = \Psi_\text{I}^{(l)} + \Psi_\text{II}^{(l)}$. In the momentum representation, with the photoelectron momentum $\vec{k} = (k,\theta,\phi)$, these partial wave functions are written as \cite{Eickhoff:2020:PRA:013430,Eickhoff:2021:PRA:052805} 
\begin{align}\label{eq:wavefunction_I_II}
    \Psi_\text{I/II}^{(l)}(\vec{k}) &=c^{(l)}_\text{I/II}(k) \, e^{i\varphi_{\mathrm{I}/\mathrm{II}}^{(l)}}  \, \psi_{3,0}(\theta,\phi)\\ \notag
    &= \vert c^{(l)}_\text{I/II}(k) \vert  \, e^{i \left( \chi^{(l)}_\text{I/II}(k) + \vphi{I/II}^{(l)}\right)}  \, \psi_{3,0}(\theta,\phi) ,
\end{align}
where $\vphi{I/II}^{(l)}$ represents the acquired optical phase and $c^{(l)}_\text{I/II}(k) = \vert c^{(l)}_\text{I/II}(k) \vert \, e^{i \chi^{(l)}_\text{I/II}(k)}$ denotes the photoelectron amplitude with the associated MPI-phase $\chi^{(l)}_\text{I/II}(k)$. Since the sum of the optical phases does not depend on the permutation of the colors, we find
\begin{equation}\label{eq:phi_I_II}
    \vphi{I}^{(l)} = 2 \vphi{g} + \varphi_{x_l} ~\text{and} ~     \vphi{II}^{(l)} =  \vphi{r} + \vphi{b} + \varphi_{x_l}
\end{equation}
for all pathways to the states $\Psi_\text{I/II}^{(l)}$. Using Eq.~\eqref{eq:wavefunction_I_II} the resulting electron density of the photoelectron interferogram $\rho^{(l)} = \vert \Psi^{(l)}\vert^2$ is given by
\begin{align}\label{eq:phot_density}
   \rho^{(l)}(\vec{k}) &= \left[ O^{(l)}(k) + S^{(l)}(k) \cos \left( \gamma^{(l)}(k) +  \Delta \varphi \right) \right] \\
   & \quad \times \,  \vert \psi_{3,0}(\theta,\phi) \vert^2, \notag
\end{align}
with the signal offset $O^{(l)}(k) = \vert c_\text{I}^{(l)}(k) \vert^2 + \vert c_\text{II}^{(l)}(k) \vert^2 $ and the signal amplitude $S^{(l)}(k) = 2 \vert c_\text{I}^{(l)}(k) \vert \vert c_\text{II}^{(l)}(k) \vert$. In Eq.~\eqref{eq:phot_density} the MPI-phase difference in channel $l$ is denoted by $\gamma^{(l)}(k) = \chi^{(l)}_\text{I}(k) -\chi^{(l)}_\text{II}(k)$ and the optical phase difference by $ \Delta \varphi=\varphi_\mathrm{I}^{(l)}-\varphi_\mathrm{II}^{(l)}$. Using Eq.~\eqref{eq:phi_I_II}, the optical phase difference reads
\begin{equation}\label{eq:phase_difference}
    \Delta \varphi = 2 \vphi{g} - \vphi{r} -\vphi{b}
\end{equation}
and is independent of the respective channel. Therefore, the phases between the interferograms are determined exclusively by the MPI-phase differences, which is the characteristic feature of our multichannel interferometry scheme. Equation~\eqref{eq:phot_density} describes the photoelectron interferograms in each channel as ${f(m=0)}$-type wave packets (cf. Fig.~\ref{fig:ExcitationScheme}(c)), which are periodically intensity-modulated by variation of the relative optical phases $\varphi_j$. The modulation frequency is determined by the number of involved photons of the respective color \cite{Eickhoff:2021:PRA:052805}. The MPI-phase differences manifest in the phase shifts $\gamma^{(l)}(k)$ of the modulation, which sensitively depend on the intermediate resonances in the MPI pathways. \\ 

\begin{table}
	\renewcommand{\arraystretch}{1.5}
	\caption{Interference channels with the relevant resonance-enhanced pathways together with the theoretical MPI-phases and the corresponding MPI-phase differences. In all cases we use $\delta_2 \geq 0$ according to the experiment.}
	\begin{tabular}{c|c|c|c|c}
		Channel &Name & Path & \parbox{1.8cm}{Theo. \\MPI-phase $\chi$\\ \quad} &\parbox{1.8cm}{Theo. MPI-phase\\difference $\gamma$\\ \quad} \\
		\hline
		\hline
		& $\mathcal{R}_\text{I}$ & ggr  & $\chi^{\text{non}}_{\delta_1<0} = \frac{3\pi}{2}$ & \multirow{-0.45}{*}{$\gamma_\text{theo}^{(1)}(\varepsilon) = \frac{ \pi}{2} - \xi(\varepsilon)$}  \\ \cline{2-4}
		\multirow{-2}{*}{$1$}
		& $\mathcal{R}_\text{II}$ & rrb  & $\chi^{\text{res}}_{\delta_1<0}(\varepsilon) =\pi + \xi(\varepsilon)$ \\ 
		\cline{1-5}
		&  $\mathcal{G}_\text{I}$ & ggg   & $\chi^{\text{non}}_{\delta_1<0} =  \frac{3 \pi}{2}$ & \multirow{-0.45}{*}{$\gamma_\text{theo}^{(2,\text{low})} = \pi$}  \\ \cline{2-4}
		&  $\mathcal{G}_\text{II}^{\text{low}}$ & brg & $\chi^{\text{non}}_{\delta_1>0} =  \frac{\pi}{2}$  \\ \cline{2-5}
		\multirow{-2}{*}{$2$}
		&  $\mathcal{G}_\text{I}$ & ggg   & $\chi^{\text{non}}_{\delta_1<0} =  \frac{3 \pi}{2}$ & \multirow{-0.45}{*}{$\gamma_\text{theo}^{(2,\text{up})} = 0$} \\ \cline{2-4}
		&  $\mathcal{G}_\text{II}^{\text{up}}$ & grb & $\chi^{\text{non}}_{\delta_1<0} =  \frac{3 \pi}{2}$  \\ 
		\cline{1-5} 
		& $\mathcal{B}_\text{I}$ & ggb & $\chi^{\text{non}}_{\delta_1<0} =  \frac{3 \pi}{2}$ & \multirow{-0.45}{*}{$\gamma_\text{theo}^{(3)} = \pi$}  \\ \cline{2-4}
		\multirow{-2}{*}{$3$}
		& $\mathcal{B}_\text{II}$ & brb & $\chi^{\text{non}}_{\delta_1>0} = \frac{\pi}{2} $  \\ \cline{2-4}
		\hline		
	\end{tabular}
	\label{tab:NotationPaths}
\end{table}
	\begin{figure*}[t]
		\includegraphics[width=0.8\linewidth]{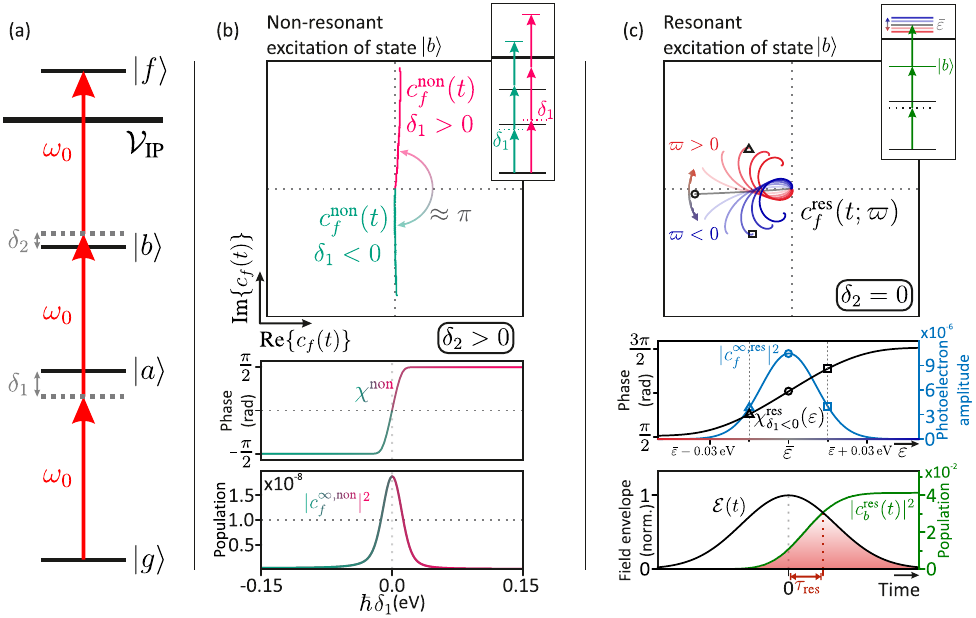}
		\caption{
			(a) Three-level system with the states $\vert g \rangle$, $\vert a \rangle$ and $\vert b \rangle$ coupled to the continuum state $\vert f \rangle$ for the excitation by a laser field with the frequency $\omega_0$ and an intensity FWHM of $\Delta t = \SI{90}{fs}$. The detunings between the exciting laser field and the respective states are $\delta_1$ and $\delta_2$. (b) Time-dependent photoelectron amplitude $c^\text{non}_f(t)$ in the complex plane representation for non-resonant excitation of the state $\vert b \rangle$ and different signs of the detuning $\delta_1$ (upper frame). Phase $\chi^\text{non} = \arg\lbrace c_f^{\infty,\text{non}} \rbrace$ and photoelectron amplitude $\vert c_f^{\infty,\text{non}} \vert^2$ as a function of the detuning $\delta_1$ (lower frames). (c) Time-dependent population amplitudes $c_f^\text{res}(t;\varpi)$ for different values of $\varpi(\varepsilon)$ assuming resonant excitation of the state $\vert b \rangle$ and $\delta_1<0$ (upper frame). The middle frame shows the photoelectron amplitude $\vert c_f^{\infty,\text{res}} \vert^2$ and corresponding MPI-phase $\chi_{\delta_1<0}^\text{res}$ as a function of $\varepsilon$. The bottom frame illustrates the physical picture of the resonance-induced photoionization time delay, explained by windowing of the temporal field envelope $\mathcal{E}$ by the time-dependent population amplitude of $\vert b \rangle$. }
		\label{fig:GenericScheme}
	\end{figure*} 
In the next step, we discuss the influence of the atomic structure on the phases $\chi^{(l)}_\text{I,II}(k)$ acquired in the MPI process by calculating the photoelectron amplitudes of the resonance-enhanced quantum pathways $\mathcal{R}_\text{I/II}$, $\mathcal{G}_\text{I/II}$ and $\mathcal{B}_\text{I/II}$ (see Tab.~\ref{tab:NotationPaths}) using perturbation theory. For simplicity, we reduce the atomic excitation scheme in Fig.~\ref{fig:ExcitationScheme}(a) to the three most relevant states, i.e., the ground state $\vert g \rangle$ and the two (near-)resonant intermediate states $\vert a \rangle$, $\vert b \rangle$ coupled to a final state $\vert f \rangle$ with variable energy in the continuum. This model system is excited by a single-color laser pulse with central frequency $\omega_0$, as shown in Fig.~\ref{fig:GenericScheme}(a). A detailed theoretical derivation of the population amplitudes is given in Appendix~\ref{app:Theo_Description}. The generic scheme describes three-photon ionization similar to that in the experiment. The MPI process is subdivided into three steps. First, the state $\vert a \rangle$ (representing the $4p$-state) is non-resonantly excited from the ground state $\vert g \rangle$ (representing the $4s$-state). For a sufficiently large detuning of the laser central frequency $\omega_0$ with respect to the transition frequency $\omega_{ga}$, i.e.,  $\delta_1 =\omega_0 - \omega_{ga} \gg \Delta \omega$, implying $\tilde{\mathcal{E}}(\delta_1) \approx 0$, the resulting time-dependent population amplitude of state $\vert a \rangle$ can be approximated by (cf. Eq.~\eqref{eq_app:a_state_off})
		\begin{equation}\label{eq:a_state_off}
		c_a^{\text{non}}(t) \propto \frac{e^{i \pi}}{\delta_1 }  e^{-i \hat{\delta}_1 t} \mathcal{E}(t).
		\end{equation}
Here $\hat{\delta}_1 = \delta_1 + \zeta$ denotes the reduced detuning which includes the phase shift $\zeta$ due to the intermediate resonance $\vert a \rangle$. This means that the time-dependent population instantaneously follows the driving electric field envelope - albeit with an additional temporal phase. For a Gaussian-shaped envelope, $\zeta$ can be estimated by (see Appendix \ref{sec_app:a_state_nonres})
\begin{equation}\label{eq:zeta}
\zeta \approx -\frac{\Delta \omega^2}{4 \ln(2) \, \delta_1}, 
\end{equation}
which explicitly depends on the detuning $\delta_1$. Consequently, the phase shift induced by the intermediate state $\vert a \rangle$, i.e. the $4p$-state, is different for the red- and blue-detuned excitation and therefore responsible for the observed energy shift in channel 2. With the experimental parameters given in Sec.~\ref{sec:phys_sys} the effective energy shift between $\varepsilon_2^\text{up}$ and $\varepsilon_2^\text{low}$ can be estimated to be about $\hbar \Delta \zeta_\text{theo} = \hbar \zeta_\text{g} - \hbar \zeta_\text{b} \approx \SI{40}{meV}$.\\

Next, we consider the two-photon excitation of state $\vert b \rangle$ (representing the $3d$-state) via the 
transiently populated off-resonant state $\vert a \rangle$ and distinguish between two limiting cases. In the two-photon resonant case, when $\delta_2= \hat{\delta}_1 + \omega_0 - \omega_{ba}=0$, the population amplitude of $\vert b \rangle$ is given by (cf. Eq.~\eqref{eq_app:b_state_integral})
    	\begin{equation}\label{eq:b_state_res}
    	c_b^{\text{res}}(t) \propto \frac{e^{-i \frac{\pi}{2}}}{\delta_1} \int_{-\infty}^{t} \mathcal{E}^2(t') \text{d}t'.  
    	\end{equation}
In contrast, for non-resonant two-photon excitation, when $\delta_2\neq0$ and the two-photon resonance lies outside the bandwidth of the second order laser spectrum implying $\tilde{\mathcal{E}}^{(2)}( \delta_2) \approx 0$, the amplitude of the state $\vert b \rangle$ reads (cf. Eq.~\eqref{eq_app:b_state_nonres})
    	\begin{equation}\label{eq:b_state_nonres}
    	c_b^{\text{non}}(t) \propto \frac{e^{-i \delta_2 t}}{\delta_1 \delta_2}  \mathcal{E}^2(t).
    	\end{equation} 
The third photon maps the population amplitude of $\vert b \rangle$ into the final continuum state $\vert f \rangle $ with photoelectron kinetic energy $\varepsilon = \hbar \omega_f - \hbar \om{IP}$. This photoionization step leads to the final state population amplitude of the free state $\vert f \rangle$ 
\begin{equation}\label{eq:photoionization}
c_f(t,\varepsilon) = - \frac{\mu_{fb}}{i \hbar} \int_{-\infty}^{t} c_b(t') \mathcal{E}(t') e^{- i \varpi(\varepsilon) t'} \text{d}t',
\end{equation}
where $\varpi(\varepsilon) = \omega_0 - \omega_{fb}(\varepsilon) = \omega_0 - \frac{\varepsilon}{\hbar}  + \omega_b - \omega_\text{IP}$. 
Photoionization via the two-photon resonant pathway yields the final ($t\rightarrow\infty$) state amplitude (cf. Eq.~\eqref{eq:app:c_f_res})
	\begin{align}\label{eq:c_state_res}
	c_f^{\infty,\text{res}}(\varepsilon) &\propto \frac{1}{  \delta_1} \bigg( \tilde{\mathcal{E}}^{(2)}(0)\tilde{\mathcal{E}}(\varpi(\varepsilon)) - 2i \mathcal{E}^2(0) \frac{\partial \tilde{\mathcal{E}} }{\partial \omega }(\varpi(\varepsilon)) \bigg) \notag \\
	&= \vert c_f^{\infty,\text{res}}(\varepsilon) \vert \,  e^{i \chi^{\text{res}}(\varepsilon)}. 
	\end{align}
In contrast, the non-resonant pathway yields the amplitude of (cf. Eq.~\eqref{eq:app:c_f_non})
	\begin{align}\label{eq:c_state_nonres}
	c_f^{\infty, \text{non}}(\varepsilon) &\propto  \frac{ e^{i \frac{\pi}{2}}}{\delta_1  \delta_2}  \tilde{\mathcal{E}}^{(3)}(\varpi(\varepsilon)+  \delta_2)\notag \\
	&= \vert c_f^{\infty, \text{non}}(\varepsilon) \vert e^{i \chi^{\text{non}}}.
	\end{align}
Equations~\eqref{eq:c_state_res} and \eqref{eq:c_state_nonres} show that for a positive detuning $\delta_2\geq0$, as in the experiment, and a symmetric and real-valued spectral envelope, the MPI-phases are determined by the sign of the detuning $\delta_1$, i.e., 
	\begin{align}
	\label{eq:varsigma_res}
	\chi^{\text{res}}(\varepsilon) &= \begin{cases}
	\pi + \xi(\varepsilon) ~;~ \delta_1 <0\\
	\xi(\varepsilon) ~;~ \delta_1 >0
	\end{cases}, \\
	\chi^{\text{non}} &= \begin{cases}
		3\pi/2 ~;~ \delta_1 <0\\
		\pi/2 ~;~ \delta_1 >0
	\end{cases}.
	\label{eq:varsigma_non}
	\end{align}
In Eq.~\eqref{eq:varsigma_res}, $\xi(\varepsilon)$ represents the temporally accumulated phase due to the resonant excitation of the $3d$-state which is discussed in the following. Since the population built-up in the resonant state occurs on the time scale of the pulse duration, the reference partial photoelectron wave packet, created by non-resonant ionization, accumulates its time-evolution phase in the continuum during the formation of the photoelectron interferogram. Thus, the time delay introduced by the population dynamics in the resonant $3d$-state manifests in a linear spectral phase between the $3d$-resonant pathway $\mathcal{R}_\text{II}$ and the non-resonant pathway $\mathcal{R}_\text{I}$. For a Gaussian-shaped pulse envelope the phase can be approximated by $\xi(\varepsilon) = \tau_\text{res} \, \varpi(\varepsilon)$ with (see Appendix \ref{app:Pert_Description} for details)
\begin{equation}\label{eq:tau_res}
\tau_\text{res} \approx \sqrt{\frac{\ln(2)}{\pi}} \frac{4}{\Delta \omega} = \frac{ \Delta t}{\sqrt{\ln(2) \pi}}  .
\end{equation}
To verify the MPI-phases predicted by the perturbation theory, we numerically solve the TDSE for the model system depicted in Fig.~\ref{fig:GenericScheme}(a). The numerical implementation is described in previous work \cite{Wollenhaupt:2005:ARPC:25,Bayer:2016:ACP:235} and the Appendix~\ref{app:Theo_Description}. The results of the simulation for the non-resonant and the resonant excitation of state $|b\rangle$ are depicted in Fig.~\ref{fig:GenericScheme} (b) and (c), respectively. In the non-resonant case, the time-dependent photoelectron amplitudes represented in the complex plane, describe almost straight lines as shown in the upper frame of (b). Each line is rotated by $\pm \pi/2$ with respect to the real axis, where the lower sign corresponds to $\delta_1 <0$ (red-detuned) and the upper sign to $\delta_1 >0$ (blue-detuned), as predicted by Eq.~\eqref{eq:varsigma_non}. This observation agrees with the variation of the MPI-phase $\chi^\text{non}$ from $\chi^{\text{non}} = -\frac{\pi}{2}$ to $\frac{\pi}{2}$ as the detuning varies from $\hbar \delta_1 = \SI{-0.15}{eV}$ to $\SI{0.15}{eV}$ as shown in the middle frame of (b). The bottom frame depicts the final population $\vert c_f^{\infty,\text{non}}\vert^2 $ as function of the detuning $\delta_1$, showing a bell-shaped profile centered around $\delta_1 \approx 0$. Note that $\delta_2$ is a function of $\delta_1$.\\

Next, we consider two-photon resonant excitation of the state $\vert b \rangle $, with $\delta_1<0$. Since resonant excitation leads to an energy-dependent MPI-phase $\chi^\text{res}_{\delta_1 <0}(\varepsilon)$ according to Eq.~\eqref{eq:varsigma_res}, we investigate the final state amplitude $c_f^\text{res}$ as a function of the photoelectron kinetic energy. The upper frame of Fig.~\ref{fig:GenericScheme}(c) shows the time evolution of  $c_f^\text{res}(t;\varpi)$ in the complex plane for different energies in the photoelectron continuum, i.e., different values of $\varpi(\varepsilon)$ (red, blue and grey lines). The red and blue shaded curves illustrate the corresponding time-dependent amplitudes for different values of $\varpi \neq 0$. The nearly horizontal grey line is connected to the final phase value of $\chi^\text{res}(\bar{\varepsilon}) = \pi$ (cf. Eq.~\eqref{eq:varsigma_res}), since the phase introduced by the population dynamics $\xi(\bar{\varepsilon})$ vanishes. In contrast, for $\varepsilon \neq \bar{\varepsilon}$, the time evolution of the energy-dependent amplitude describes a curve moving either upwards ($\varpi >0$, red lines) or downwards ($\varpi <0$, blue lines) on the imaginary axis. The middle frame of (c) depicts the energy-resolved photoelectron amplitude $\vert c_f^{\infty,\text{res}} \vert^2$ (blue) and phase $\chi^\text{res}_{\delta_1 <0}(\varepsilon)$ (black). Here, the symbols (circle, square and triangle) indicate the endpoints of the final state populations shown in the upper frame. In the region of a non-vanishing photoelectron amplitude $\vert c_f^{\infty,\text{res}} \vert^2$, the phase exhibits a linear behavior, confirming the assumption of $\xi(\varepsilon) \approx \tau_\text{res} \varpi(\varepsilon)$. The bottom frame shows a physical picture of the resonance-induced ionization time delay due to the windowing of the Gaussian-shaped temporal field envelope $\mathcal{E}(t)$ with the time-dependent population amplitude of the state $\vert b \rangle$. While non-resonant excitation and ionization follows the field envelope almost instantaneously (cf. Eq.~\eqref{eq:a_state_off}), resonant ionization introduces a temporal shift in the order of the pulse duration according to Eq.~\eqref{eq:tau_res}, corresponding to a linear spectral phase. 
	
\section{Experiment}	
\label{sec:experiment}

In this section we describe our experimental setup and strategy. The trichromatic fields are generated using white light polarization pulse shaping \cite{Kerbstadt:2017:JMO:1010,Kerbstadt:2017:OE:12518,Eickhoff:2021:PRA:052805} combined with photoelectron velocity map imaging (VMI) \cite{Eppink:1997:RSI:3477} to measure energy- and angle-resolved photoelectron spectra.

\subsection{Experimental setup}\label{sec:setup}
\begin{figure}[t]
	\includegraphics[width=\linewidth]{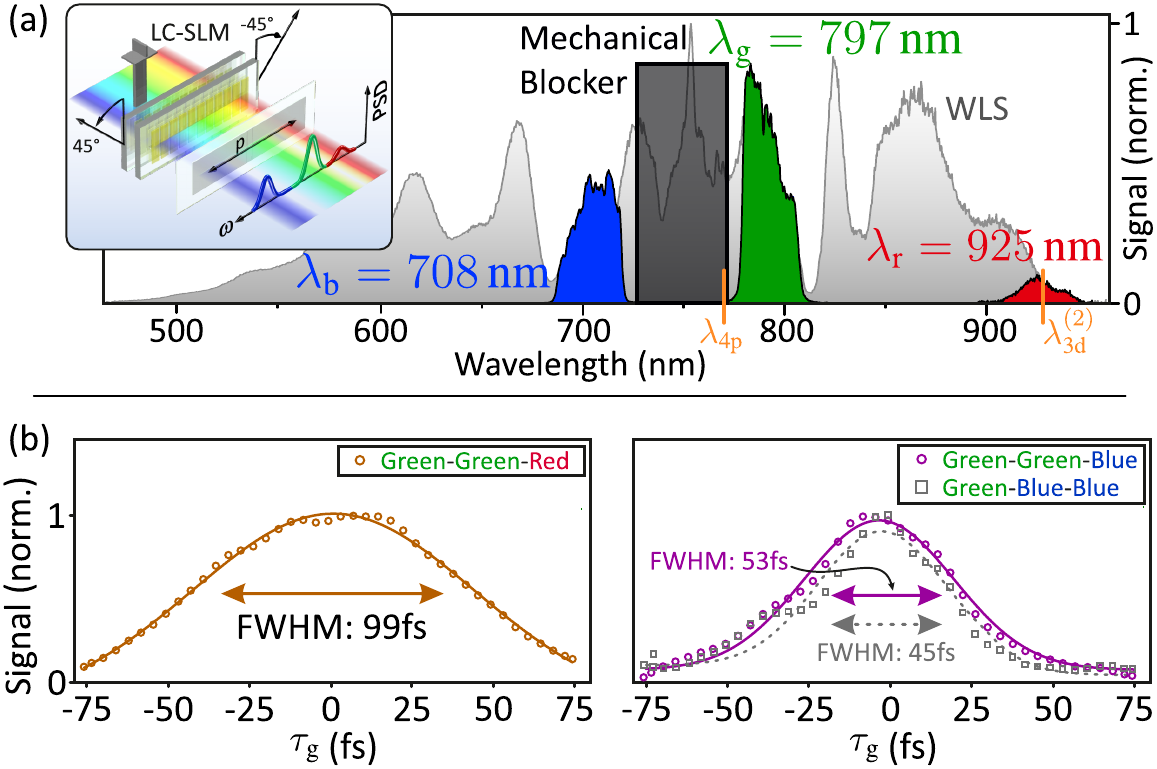}
	\caption{
	(a) Measured trichromatic spectrum together with the WLS. The inset shows a schematic of the pulse shaper setup with a broadband $p$-polarizer and the mechanical blocker in the Fourier-plane to eliminate resonant excitation of the $4p$-state. (b) Energy sections of the time-resolved measurements on the \textit{green-red} (left) and \textit{green-blue} (right) frequency mixing signals from \textit{in-situ} temporal pulse characterization (circles and squares) together with Gaussian-fits (solid and dashed lines) and corresponding FWHMs.
}
	\label{fig:Strategy}
\end{figure}

The trichromatic experimental scheme is based on our setup previously reported in \cite{Eickhoff:2021:PRA:052805}. Briefly, trichromatic pulse shaping is implemented using a home-built $4f$ polarization pulse shaper \cite{Kerbstadt:2017:JMO:1010} employing a dual-layer liquid crystal spatial light modulator (LC-SLM). The shaper is specifically adapted to the white light supercontinuum (WLS) \cite{Kerbstadt:2017:OE:12518} generated by seeding a neon-filled hollow-core fiber compressor with \SI{20}{fs} pulses ($\lambda_c = \SI{790}{nm}$) from an amplified laser system (\textsc{Femtolasers} Femtopower HR \SI{3}{kHz}). Employing a custom broadband $p$-polarizer (\textsc{Codixx} colorPol$^{\tiny{\text{\circledR}}}$) in the Fourier plane of the shaper, we generate a PLP trichromatic field (cf. inset in Fig.~\ref{fig:Strategy}(a)) by spectral amplitude and phase modulation. To individually advance or delay the pulses of different colors in time, we apply linear spectral phase functions to the LC-SLM. Owing to the common-path geometry of the shaper-based scheme, the generated trichromatic pulse sequences are inherently phase-locked allowing for high-precision time- and phase-resolved measurements \cite{Koehler:2011:OE:11638}. The measured spectrum of the trichromatic field is shown in Fig.~\ref{fig:Strategy}(a) along with the input WLS. The central wavelength of the red pulse is set to $\lambda_{\text{r}}=\SI{925}{nm}$ (red), i.e. tuned to the two-photon resonance of the $3d$-state, indicated by the right vertical orange line. The central wavelengths of the green and blue pulses are set to $\lambda_{\text{g}}=\SI{797}{nm}$ and $\lambda_{\text{b}}=\SI{708}{nm}$, respectively. Spectral components resonant with the $4p$-state are eliminated by employing a mechanical blocker (cf. inset in Fig.~\ref{fig:Strategy}(a)) in the Fourier-plane between the blue and green component (see also left vertical orange line). The spectral bandwidths of the red, green and blue bands are set to $\Delta \om{r} = \SI{50}{mrad/fs}$, $ \Delta \om{g} = \SI{90}{mrad/fs}$ and $ \Delta \om{b} = \SI{100}{mrad/fs}$ (FWHM of the intensity), corresponding to bandwidth-limited pulse durations of $\Delta t_{\text{r}} \approx \SI{55}{fs}$, $\Delta t_{\text{g}} \approx \SI{32}{fs}$ and $\Delta t_{\text{b}} \approx \SI{28}{fs}$, respectively, assuming Gaussian-shaped envelopes. Using a spherical mirror, the trichromatic pulse is focused into the interaction region of a VMI spectrometer filled with potassium vapor from a dispenser source. The peak intensity of the temporally overlapping trichromatic field in the interaction region is estimated to be in the order of $I_0 \approx \SI{5e12}{W/cm^2}$. Abel inversion of the recorded 2D VMI images, using the \textsc{pBasex} algorithm \cite{Garcia:2004:RSI:4989}, yields the PMD of the created photoelectron wave packets. Angle-integration and energy calibration \cite{Wituschek:2016:RSI:083105} of the PMD leads to the energy-resolved photoelectron spectra shown in Fig.~\ref{fig:ExcitationScheme}(b). 

\subsection{Experimental preparation and characterization}\label{sec:exp_strategy}
In this section, we describe how we prepare and characterize the shaper-generated trichromatic field for the experiment. To maximize the interference contrast in all three interference channels $l$ (cf. Fig.~\ref{fig:ExcitationScheme}(a)), we optimize the temporal overlap and the amplitudes of the three colors together with the spectral overlap of the partial photoelectron wave packets in each channel. First, to compress the WLS in time, we use a shaper-based evolutionary optimization procedure to maximize the second harmonic yield of the WLS from a thin $\beta$-barium borate crystal (\textsc{Eksma Optics}, cutting angle $\theta = \SI{29.2}{\degree}$, $\SI{5}{\micro m}$ thickness) \cite{Baumert:1997:APB:779,Yelin:1997:OL:1793}. Subsequently, an initial trichromatic spectrum is sculpted from the compressed WLS by spectral amplitude modulation. In the second step, the frequency mixing contributions in the second harmonic signal of the trichromatic field are maximized by shaper-based pairwise optimization of the second and third order dispersion. The refined and temporally compressed trichromatic field is then focused into the interaction region of the VMI spectrometer to perform an additional \textit{in-situ} temporal characterization and an iterative spectral optimization of the sculpted field in the laser focus. To this end, we iteratively adjust the amplitude and central wavelength of the individual colors to maximize the contrast of the photoelectron interferograms in phase maps \cite{Eickhoff:2021:PRA:052805}. For the final characterization of the temporal overlap we perform bichromatic time-resolved measurements by varying the time delay $\ta{g}$ of the green pulse with a stepsize of $\Delta \tau_\mathrm{g} = \SI{3.75}{fs}$ and record the \textit{Green-Red} and the \textit{Green-Blue} photoelectron frequency mixing contributions. Figure~\ref{fig:Strategy}(b) shows sections from these time-resolved measurements at kinetic energies $\varepsilon_1$ (left) as well as $\varepsilon_3$ and $\varepsilon_4$ (right) together with Gaussian fits (solid and dashed lines). The results show a maximal temporal overlap of the green and red pulse at $ \SI{1.7(08)}{fs}$ and of the green and blue pulse at 
$\SI{-3.5(13)}{fs}$, i.e. within the temporal step size, ensuring efficient frequency mixing. By applying the procedure described in \cite{Eickhoff:2021:PRA:052805} we obtain pulse durations of $\Delta t_\text{r} = \SI{50(8)}{fs}$, $\Delta t_\text{g} = \SI{49(12)}{fs}$ and $ \Delta t_\text{b} = \SI{29(12)}{fs}$ from the measured FWHMs shown in Fig.~\ref{fig:Strategy}(b). The values for the red and blue pulse are in good agreement with the estimated bandwidth-limited pulse durations from Sec.~\ref{sec:setup}. The retrieved pulse duration of the green component is slightly larger than the value obtained in Sec.~\ref{sec:setup}, which is attributed to the mechanically blocked edge of the green spectral component resulting in a non-Gaussian pulse shape. Overall, the results confirm a flat spectral phase of the trichromatic field. 

\subsection{Experimental strategy}\label{sec:exp_data_evaluation}
To extract the MPI-phase differences from the measured photoelectron interferograms, the experiment consists of two parts. In Sec.~\ref{sec:OptControlPhases}, we describe the energy- and phase-resolved measurements, i.e., phase maps. In the experiment, we vary the relative optical phase of the green pulse $\vphi{g}$ and measure photoelectron spectra which are energy-calibrated \cite{Wituschek:2016:RSI:083105} and angularly integrated
\begin{equation}
\varrho_\text{int}(\varepsilon) = \int_{0}^{2 \pi} \varrho(\varepsilon,\theta) \text{d}\theta,
\end{equation}
yielding the kinetic energy-resolved phase maps $\varrho_\text{int}(\varepsilon, \vphi{g};\vphi{r} + \vphi{b})$, with $\varepsilon = \frac{(\hbar k)^2}{2m_e}$. 
Changing the phase of the blue pulse by $\pi$ inverts the measured phase maps. In Sec.~\ref{sec:InducedPhases} this inversion is used to remove the phase-insensitive background. To this end, we calculate the phase contrast map \cite{Eickhoff:2021:PRA:052805}
\begin{align}\label{eq:asymm}
\mathcal{C}(\varepsilon,\vphi{g}) &=\varrho_{\text{int}}(\varepsilon,\vphi{g};0) -  \varrho_{\text{int}}(\varepsilon,\vphi{g};\pi) \notag \\
&\propto \cos \left( \gamma^{(l)}(\varepsilon) +  2\vphi{g} \right),
\end{align}	
by subtracting two normalized measured phase maps with an effective phase difference of $\vert \vphi{r} + \vphi{b}\vert = \pi$. The phase contrast map reveals even subtle imprints of the MPI-phases in the recorded photoelectron spectra. In particular, the splitting of channel 2 into an upper and a lower contribution corresponding to the red- and blue-detuned excitation of the $4p$-state can be clearly seen. In addition, it directly maps the spectral phase due to the resonant excitation of the $3d$-state in the interference channel~1.\\

\section{Results and discussion}\label{sec:results_discussion}
In Sec.~\ref{sec:OptControlPhases}, we present the $\vphi{g}$-resolved phase maps and demonstrate their inversion by changing optical phase of the blue pulse from  $\vphi{b}=0$ to $-\pi$. In Sec.~\ref{sec:InducedPhases}, we remove the phase-insensitive background by calculating the phase contrast map $\mathcal{C}(\varepsilon,\vphi{g})$ (cf. Eq.~\eqref{eq:asymm}) and retrieve the MPI-phases induced by the intermediate states $4p$ and $3d$.

\subsection{Phase control of the photoelectron interferograms}\label{sec:OptControlPhases}
\begin{figure}[t]
	\includegraphics[width=\linewidth]{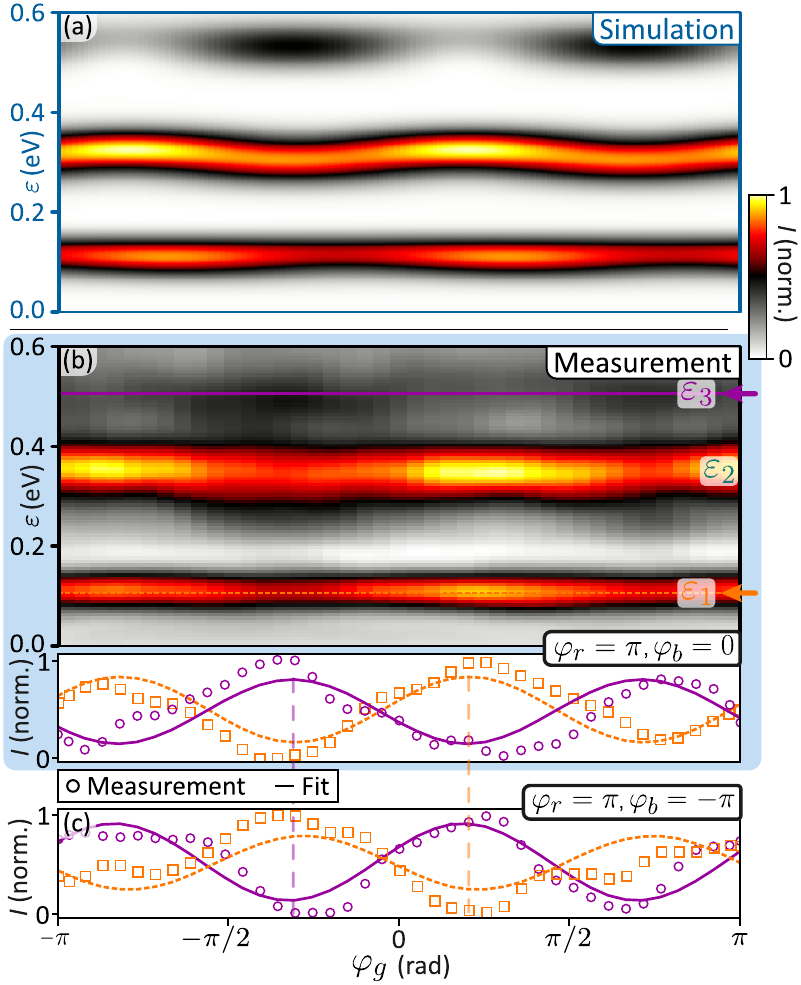}
	\caption{
	Simulated and measured phase maps of photoelectron interferograms from multichannel interferometry by varying the phase of the green pulse $\vphi{g} \in [-\pi , \pi]$. (a) Simulated phase map obtained by numerical solution of the TDSE for the perturbative interaction of a three-level atom with the trichromatic field having optical phases of $\vphi{r}=\pi$, $\vphi{b}=0$. (b) Measured phase map for $\vphi{r}=\pi$, $\vphi{b}=0$ with a step size of $\delta \vphi{g} = \frac{2\pi}{40} \approx \SI{0.16}{rad}$, showing the three energetically separated interference channels around $\varepsilon_1$-$\varepsilon_3$. Bottom frame: section through the phase maps along the phase axis at $\varepsilon_1$ (orange) and $\varepsilon_3$ (violet) to emphasize the modulation. (c) same as the bottom frame of (b) but for $\vphi{r}=\pi$, $\vphi{b}=-\pi$.}
	\label{fig:Results}
\end{figure}

First, we investigate the phase control of the photoelectron interferograms in $\vphi{g}$-resolved phase maps $\varrho_\text{int}(\varepsilon, \vphi{g};\vphi{r} + \vphi{b})$. In Fig.~\ref{fig:Results} (a) and (b) the simulated and measured phase maps $\varrho_\text{int}(\varepsilon, \vphi{g};\pi)$ are compared. The observed phase dependence validates the scheme by showing that the same optical phase is imprinted in all interferograms (cf. Eq.~\eqref{eq:phot_density}). \\

In the experiment, we vary $\vphi{g} \in [-\pi , \pi]$ with a step size of $\delta \vphi{g} = \frac{2\pi}{40}$ and set the phases of the red and blue pulse to $\vphi{r}=\pi$ and $\vphi{b}=0$, respectively. The upper frame of Fig.~\ref{fig:Results}(b) depicts the measured phase map which shows the three energetically separated interference channels centered around $\varepsilon_1$, $\varepsilon_2$ and $\varepsilon_3$ in an energy-window of $\varepsilon \in [0, 0.6]\,\text{eV}$. In accordance with the theoretical description in Sec.~\ref{sec:pert_description}, each channel is strongly amplitude modulated with a periodicity of $\pi$, however, each with an individual phase offsets due to MPI-phases. The $\pi$-periodicity is due to the fact, that effectively two green photons contribute to each interferogram, resulting in a modulation period of $\frac{2\pi}{2} = \pi$ (cf. Eq.~\eqref{eq:phase_difference}). The results in channel 2 show small energetic oscillations of the signal attributed to the two slightly shifted contributions discussed in Sec.~\ref{sec:phys_sys}. These two contributions, which are hardly distinguishable at this point, are more clearly visible in the phase contrast map and, therefore, they will be discussed in detail in Sec.~\ref{sec:InducedPhases}. First, we analyze the modulations in channels 1 and 3 around $\varepsilon_1$ and $\varepsilon_3$. 

	\begin{figure*}[t]
		\includegraphics[width=\linewidth]{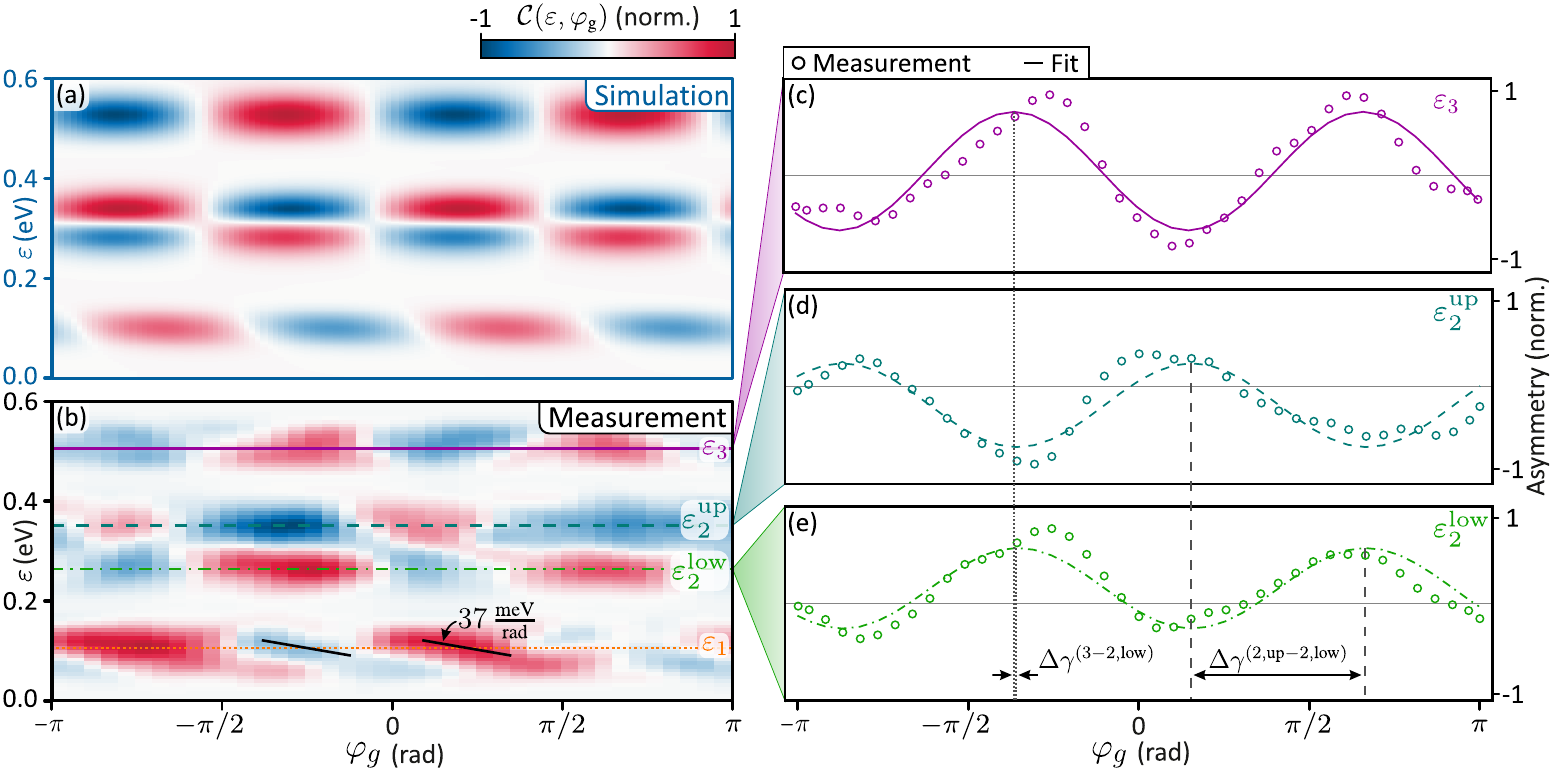}
		\caption{
			Simulated (a) and measured (b) phase contrast maps $\mathcal{C}(\varepsilon,\vphi{g})$ with $\vphi{r}=\pi$ extracted from inverting the phase maps by changing the phase of the blue pulse from $\vphi{b}=0$ to $\vphi{b}=-\pi$. (a) Simulated phase contrast map based on the numerical results described in Sec.~\ref{sec:OptControlPhases}, shown in Fig.~\ref{fig:Results}(a). (b) Measured phase contrast map calculated from the results of Sec.~\ref{sec:OptControlPhases}, shown in Fig.~\ref{fig:Results}(b).The different channels are indicated by horizontal lines. To emphasize the interference channels, we apply super-Gaussian filters to blank the space between the regions of interest. (c)-(e) Sections through the measured phase contrast map (circles) at the kinetic energies of $\varepsilon_3$ (c), $\varepsilon_2^\text{up}$ (d) and $\varepsilon_2^\text{low}$ (e) together with cosine fits through the data (lines). The MPI-phase differences $\Delta \gamma$ are highlighted by vertical dashed black lines through the frames (c)-(e).
		}
		\label{fig:Asymm}
	\end{figure*}
To quantitatively evaluate the modulation of the photoelectrons in the channels 1 and 3, we take sections through the phase maps along the phase axis at $\varepsilon_1$ (orange squares) and $\varepsilon_3$ (violet circles). These sections are shown in the bottom frame of Fig.~\ref{fig:Results}(b) together with cosine fits through the data (orange dashed and violet solid lines). The inversion of the phase map is demonstrated by setting $\vphi{b}=-\pi$. The experimental result of the inversion is shown in Fig.~\ref{fig:Results}(c). Comparison of the orange (channel 1) and violet lines (channel 3) yields a phase shift of $\SI{1.6}{rad} \approx \pi/2$, which confirms the inversion of the interference patterns. Minor imperfections of the inverted phase map are attributed to diffraction on the discrete phase mask of the SLM \cite{Weiner:2000:RSI:1929,Meshulach:1998:Nature:239}. The phase shifts between the modulations in all channels will be discussed in Sec.~\ref{sec:InducedPhases}.\\

Our interpretation of the experimental findings are confirmed by the simulated phase map in Fig.~\ref{fig:Results}(a) for $\vphi{r}=\pi$, $\vphi{b}=0$ which is in good agreement with the measured phase map in the upper frame of (b). In the simulation, we consider a three-level system consisting of the states $4s$, $4p$ and $3d$ (similar to Sec.~\ref{sec:pert_description} and the Appendix \ref{app:Theo_Description}) which interacts with a sequence of three Gaussian-shaped pulses similar to those used in the experiment (see Sec.~\ref{sec:experiment}). By numerically solving the TDSE for the three-level system, we obtain the bound state population dynamics. Subsequently, we calculate the energy-resolved photoelectron spectra using time-dependent perturbation theory (see e.g. \cite{Wollenhaupt:2006:PRA:063409} for details). 

\subsection{MPI-phase retrieval}
\label{sec:InducedPhases}
	
In this section, we study the phase shifts between the interferograms observed in the different interference channels. These phase shifts originate from the MPI-phases acquired in the corresponding photoionization processes. For the background-free determination of the phase shifts in Fig.~\ref{fig:Results}, we calculate the phase contrast maps $\mathcal{C}(\varepsilon,\vphi{g})$ according to Eq.~\eqref{eq:asymm} from the phase maps presented in Sec.~\ref{sec:OptControlPhases}. The simulated and experimental phase contrast maps are depicted in Fig.~\ref{fig:Asymm}(a) and (b), respectively. To highlight the measured interferograms in Fig.~\ref{fig:Asymm}(b), we apply super-Gaussian filters to blank the space between the regions of interest. The phase contrast map reveals the splitting of channel 2 which is barely visible in the phase map shown in Fig.~\ref{fig:Results}(b). The two contributions are indicated by the petrol-blue dashed and green dash-dotted lines at $\varepsilon_2^\text{up}  = \SI{0.35}{eV}$ and $\varepsilon_2^\text{low} = \SI{0.28}{eV} $, respectively. In this representation, the maxima have a spectral separation of about $\hbar \, \Delta \zeta_\text{meas} \approx \SI{70}{\milli\electronvolt}$ which exceeds the theoretically expected separation of $\hbar \Delta \zeta_\text{theo} \approx \SI{40}{meV}$ from Eq.~\eqref{eq:zeta} (cf. Sec.~\ref{sec:pert_description}). The deviation is explained by the near-resonant excitation $\delta_1 \approx \Delta \omega$ in the experiment, in contrast to the assumption of $\delta_1 \gg \Delta \omega$ in the theoretical description. However, since the mechanical blocker in the Fourier-plane produces a strongly non-Gaussian optical spectrum with vanishing intensity at resonance, the assumption $\tilde{\mathcal{E}}(\delta_1) \approx 0$ is still valid. Moreover, the choice of frequencies in the experiment leads to a slight energetic separation in the frequency-mixing and the green single-color signals of \SI{20}{meV}. The results show that both contributions in channel 2 exhibit a $\pi$-periodic oscillation similar to those observed in channel 1 and 3 (see discussion in Sec.~\ref{sec:OptControlPhases}). Again, for quantitative evaluation of the MPI-phases, we take energy sections through the phase contrast map along the $\vphi{g}$-axis. The sections at $\varepsilon_3$ (violet circles), $\varepsilon_2^\text{up}$ (petrol blue circles) and $\varepsilon_2^\text{low}$ (green circles) are shown in Fig.~\ref{fig:Asymm}(c)-(e) along with cosine fits through the data (lines). The phases $\gamma^{(l)}$ extracted from the channels ${l = 2(\text{up/low}),3}$ provide the MPI-phase difference  
\begin{equation}
	\Delta \gamma^{(l-l')} = \gamma^{(l)}- \gamma^{(l')}
\end{equation}
between two channels $l$ and $l'$. To rationalize the extracted phase differences $\Delta \gamma$, we discuss the values $\Delta \gamma^{(3-2,\text{low})}$ and $\Delta \gamma^{(2,\text{up}-2,\text{low})}$ by comparing them with the theoretical results, starting with $\Delta \gamma^{(3-2,\text{low})}$. In both the channels 3 and (2, low), two ionization pathways interfere, one of which originates from a blue-detuned near-resonant transition, the other from a red-detuned near-resonant transition. Consequently, the MPI-phase difference between the interference channel 3 and the lower part of channel 2 is expected to vanish (cf. Tab.~\ref{tab:NotationPaths}), in good agreement with the measured phase difference of $\Delta \gamma^{(3-2,\text{low})} = \SI{0.0}{rad}$.\\

In multichromatic photoionization, different final states are addressed with different combinations and permutations of the colors in the laser field. However, the pathways to create the photoelectrons in the interference channel 2, i.e., $\mathcal{G}_\text{II}^\text{up}$ at $\varepsilon_2^\text{up}$ and $\mathcal{G}_\text{II}^\text{low}$ at $\varepsilon_2^\text{low}$ differ only in the permutations of the photons -- albeit with the common reference pathway $\mathcal{G}_\text{I}$ (cf. Sec.~\ref{sec:phys_sys} and Fig.~\ref{fig:ExcitationScheme}(a)). Therefore, the phase difference within the interference channel 2 is uniquely determined by the MPI-phases introduced by the red-detuned (upper signal) and blue-detuned (lower signal) (1+2) REMPI processes. To analyze the dynamics, we first note that the red/blue detuning of the field corresponds to an opposite sign of $\delta_1$. As a consequence, the two contributions in the interference channel 2 map the transient population dynamics of the $4p$-state but with opposite signs in the amplitude $ 1/\delta_1$ and in the time-dependent phase $\hat{\delta}_1 t$ as described by Eq.~\eqref{eq:a_state_off}. While the sign change in the phase explains the previously described energetic splitting, the different signs of the amplitudes lead to a theoretical MPI-phase difference of $\Delta \gamma_\text{theo}^{(2,\text{up}-2,\text{low})} = - \pi$ (cf. Fig.~\ref{fig:GenericScheme}(b) and Tab.~\ref{tab:NotationPaths}). The measured MPI-phase difference of $\Delta \gamma^{(2,\text{up}-2,\text{low})} = \SI{-3.1}{rad}$ confirms the above physical picture.\\

Finally, we discuss the interference pattern observed in channel 1 by analyzing the phase contrast map shown in Fig.~\ref{fig:Asymm}. Both, the measured and the simulated phase contrast maps in frame (a) and (b) show a pronounced spectral shearing around $\varepsilon_1$ corresponding to an approximately linear spectral phase. Following the discussion in Sec.~\ref{sec:pert_description}, we explain the linear spectral phase by identifying a time delay in the resonant photoionization. The (2+1) REMPI pathway $\mathcal{R}_\text{II}$ in channel 1 consists of the 2-photon resonant excitation of the $3d$-state with the red pulse and the 1-photon ionization with the blue pulse. The physical picture to rationalize the time delay is illustrated in the bottom frame of Fig.~\ref{fig:GenericScheme}(c). Due to the 2-photon resonance, the time evolution of the population amplitude of the $3d$-state has a sigmoidal shape. Since the REMPI process is characterized by the ionization of the resonant $3d$-state, the product of the temporal envelope of the ionizing pulse and the time-dependent population amplitude appear in the perturbative description of the photoionization in Eq.~\eqref{eq:photoionization}. Hence, the temporal envelope of the ionizing blue pulse $\mathcal{E}_\text{b}(t)$ is windowed by the sigmoidal-shaped population amplitude $c_b^{\text{res}}(t)$. The windowing results in a time delayed formation of the photoelectron wave packet from (2+1) REMPI with respect to the temporal pulse envelope. The interference of the time delayed wave packet from the resonant (2+1) REMPI pathway $\mathcal{R}_\text{II}$ with the wave packet from the near-resonant (1+2) REMPI pathway $\mathcal{R}_\text{I}$ in channel 1 gives rise to the spectral shearing in the phase map \cite{Eickhoff:2021:PRA:052805,Pengel:2017:PRL:053003,Pengel:2017:PRA:043426}. Assuming a Gaussian-shaped temporal pulse envelope $\mathcal{E}(t)$, the associated linear spectral phase can be approximated around the center of the signal ($\varpi(\varepsilon)=0$) by $\xi(\varepsilon) \approx - \frac{\ta{res}}{\hbar} \, \varepsilon + \xi_0$ with the resonance-induced time delay $\ta{res}$ from Eq.~\eqref{eq:tau_res} (see Appendix \ref{sec:app:Res_b} for a detailed discussion). For the field parameters used in the experiment (Sec.~\ref{sec:exp_strategy}) we calculate a delay of $\ta{res} \approx \SI{19}{fs}$ which corresponds to a slope of $\SI{37}{meV/rad}$. This slope derived from theory is indicated in Fig.~\ref{fig:Asymm}(b) and in good agreement with the observed shearing in the phase contrast map of channel 1. Our results show that multichannel interferometry with temporally overlapping trichromatic fields allows to unambiguously determine the MPI-phases acquired in REMPI processes.

\section{Summary and conclusion}

In this work, we presented a pulse shaper-based trichromatic multichannel wave packet interferometry scheme to study the quantum phases induced by multiphoton ionization (MPI) dynamics. We demonstrated the scheme on the MPI of potassium atoms with temporally overlapping parallel linearly polarized trichromatic fields. By appropriate choice of the optical frequencies so that $2\omega_\mathrm{g} \approx \omega_\mathrm{r}+\omega_\mathrm{b} < \omega_\text{IP}$, the photoelectron wave packets overlap in different interference channels and create three energetically separated interferograms. These interferograms are sensitive to both the optical and the MPI-phases acquired along the different MPI pathways. Due to the design of the scheme, the optical phase is identical in all interferograms. Therefore, the measured phase shifts directly revealed the MPI-phases introduced by the MPI dynamics in the different interference channels. Specifically, the red pulse was tuned to the two-photon $4s\rightarrow\rightarrow3d$ transition, opening a (2+1) REMPI pathway via ionization with the blue pulse. The green and the blue pulse were tuned close to the $4p$-resonance with opposite detunings to implement two different (1+2) REMPI pathways via trichromatic frequency mixing. We observed high-contrast phase-shifted intensity modulations in the photoelectron interferograms by varying the green pulse's relative phase. By introducing an optical phase of $-\pi$ to the blue pulse, the interferograms were inverted and used to remove the phase-insensitive background. Analysis of the background-free interferograms unambiguously revealed the MPI-phases introduced by the intermediate resonances.\\

The measured interferograms showed that the MPI-phases in the photoelectron interferogram manifest in three different ways. First, we observed an energetic splitting of the photoelectron interferograms from red- and blue-detuned (1+2) REMPI. Although interferograms from frequency mixing pathways consisting of the permutations of different should be energetically degenerate, we demonstrated that the near- but non-resonant excitation of the intermediate $4p$-state lifts this degeneracy. Second, we observed constant phase shifts of $\pi$ both between the split interferograms within one interference channel and between interferograms of different interference channels. These phase shifts between the different (1+2) REMPI signals originate from population amplitudes with opposite signs. Finally, we demonstrated that the photoionization time delay through the resonant excitation of the $3d$-state in the (2+1) REMPI pathway gives rise to a linear spectral phase in the photoelectron interferograms observed in the pronounced spectral shearing. All observations were analyzed using an analytic perturbative description and numerical simulations of a trichromatic driven three-level system coupled to the continuum.\\

In conclusion, we demonstrated that our shaper-based trichromatic multichannel interferometry scheme is a powerful tool to measure the phase of the ultrafast free electron wave packets. We showed that the resonance-induced photoionization time delay associated with a linear spectral phase is a general feature of processes involving a resonant intermediate state. For example, the structure of bichromatic photoelectron vortices as reported in \cite{Kerbstadt:2019:NC:658,Kerbstadt:2019:APX:1672583,Eickhoff:2021:JPBAMOP:164002} is affected by the intermediate resonances. Currently, we investigate further applications of multichromatic free electron vortex spectroscopy in our labs. In general, our results show that multicolor schemes open up entirely new perspectives for the measurement and coherent control of ultrafast quantum dynamics if we succeed in harnessing the great variety of interference pathways by a suitable design of the experiment.

\section{Acknowledgments}
Financial support by the Deutsche Forschungsgemeinschaft via the priority programme SPP1840 QUTIF is gratefully acknowledged.

\appendix

\section{Generic three-level system}
\label{app:Theo_Description}

For the convenience of the reader and to introduce our notation, this appendix provides a detailed description of the multichannel interferometry scheme studied in the experiment. In this section, we describe the perturbative excitation of a generic three-level atom by a laser electric field $E(t) = \mathcal{E}(t) e^{-i \omega_0 t}$, with the pulse envelope $\mathcal{E}(t)$ and the central frequency $\omega_0$. The three-level model system, as depicted in Fig.~\ref{fig:GenericScheme}(a), consists of the states $\vert g \rangle$, $\vert a \rangle$ and $\vert b \rangle$ with the corresponding transition frequencies $\omega_{ag} = \omega_a - \omega_g $ and $\omega_{ba} = \omega_b - \omega_a $ and non-zero dipole couplings $\mu_{ag}=\langle a|\mu|g\rangle$ and $\mu_{ba}=\langle b|\mu|a\rangle$. Expanding the system state into this basis, i.e. $|\psi(t)\rangle=\sum_{n=g,a,b} c_n(t)|n\rangle$, the laser-atom interaction is described by the TDSE
\begin{equation}
i \hbar \frac{d}{dt} \begin{pmatrix}
c_g(t)\\ 
c_a(t)\\ 
c_b(t)
\end{pmatrix} =  \mathcal{H}(t) \begin{pmatrix}
c_g(t)\\ 
c_a(t)\\ 
c_b(t)
\end{pmatrix}
\end{equation}
for the population amplitudes $c_n(t)$. The Hamiltonian is given in the dipole approximation by
\begin{equation}
\mathcal{H}(t)=  \left(\begin{array}{ccc}
0  & -\mu_{ga}E^*(t)  & 0 \\
-\mu_{ag}E(t)	 & \hbar\omega_a  & -\mu_{ab}E^*(t) \\
0  & -\mu_{ba}E(t)  & \hbar\omega_b
\end{array}\right).
\end{equation}
The TDSE is solved iteratively on a discrete temporal grid with stepsize $\delta t$ using the short-time propagator technique \cite{Feit:1974:APL:169,Cohen-Tannoudji:1977:3,Wollenhaupt:2010:PRA:053422}
\begin{equation}
\begin{pmatrix}
	c_g(t+\delta t)\\ 
	c_a(t+\delta t)\\ 
	c_b(t+\delta t)
\end{pmatrix}
=e^{-i\frac{\delta t}{\hbar}\mathcal{H}(t)}\cdot
\begin{pmatrix}
	c_g(t)\\ 
	c_a(t)\\ 
	c_b(t)
\end{pmatrix}
\end{equation}
and assuming the atom to be initially in the ground state. The photoionization of the population in state $\vert b \rangle$ is described by the excitation of a final free electron state $\vert f \rangle$ with $\omega_{fb}(\varepsilon) = \omega_f(\varepsilon) - \omega_b $ as shown in Fig.~\ref{fig:GenericScheme}(a).

\section{Perturbative description of MPI-phases}
\label{app:Pert_Description}

\subsection{Excitation of the state $\vert a \rangle$}\label{sec_app:state_a_excitation}
The population amplitude $c_a(t)$ of the state $\vert a \rangle$ is given by $1^\text{st}$ order perturbation theory \cite{Meshulach:1999:PRA:1287,Cohen-Tannoudji:2019} 
\begin{align}\label{eq:c_a_general}
c_a(t) &= - \frac{\mu_{ag}}{i \hbar} \int_{- \infty}^{t} \mathcal{E}(t') e^{-i \delta_1 t'} \text{d}t',
\end{align}
with the detuning $\delta_1 = \omega_0 - \omega_{ag} $ and the transition dipole moment $\mu_{ag}$. For $t \rightarrow \infty$, Eq.~\eqref{eq:c_a_general} leads to the well-known result $c_a(\infty) = - \frac{\mu_{ag}}{i \hbar} \tilde{\mathcal{E}}(\delta_1)$. Hence, the final population amplitude of state $\vert  a \rangle$ is given by the spectral amplitude at the transition frequency. In general, by extending the integral in Eq.~\eqref{eq:c_a_general} up to infinity, utilizing the Heaviside function $\theta(t)$, we obtain
\begin{align}
c_a(t) &= - \frac{\mu_{ag}}{i \hbar} \int_{- \infty}^{\infty} \mathcal{E}(t') e^{-i \delta_1 t'} \theta(t-t')\text{d}t'.
\end{align}
Applying the Fourier transform in combination with the convolution theorem\footnote{Here, we use $
	\mathcal{F}\lbrace f \rbrace  (\omega) \define  \int_{- \infty}^{\infty} f(t) e^{- i \omega t} \text{d}t
	$ as the convention for the Fourier transform of the function $f$ and the convolution theorem $\mathcal{F}\lbrace f \cdot g \rbrace (\omega) = \frac{1}{2\pi} 	\mathcal{F}\lbrace f \rbrace (\omega) \otimes \mathcal{F}\lbrace  g \rbrace (\omega)$.}
the amplitude can be written as	
\begin{align}
c_a(t) &= - \frac{\mu_{ag}}{2 \pi i \hbar} \bigg( \bigg[ \pi \delta(\omega) + \frac{i e^{- i \omega t}}{\omega} \bigg]  \otimes \tilde{\mathcal{E}}(\omega)  \bigg) (\delta_1).
\end{align}
Note that the convolution integral of the second term is only defined as its Cauchy principal value \cite{Dudovich:2002:PRL:123004,Osgood:2019}, denoted as $\fint$. Hence, the time-dependent population amplitude of $\vert a  \rangle$ reads
\begin{equation}\label{eq_app:a_state_full}
c_a(t)= - \frac{\mu_{ag}}{2 \pi i \hbar} \bigg( \pi \tilde{\mathcal{E}}(\delta_1) - i  \fint_{-\infty}^{\infty} \frac{\tilde{\mathcal{E}}(\omega) e^{i (\omega- \delta_1) t}}{ \omega - \delta_1} \text{d}\omega    \bigg).
\end{equation}

\subsubsection{Resonant excitation of the state $\vert a \rangle$: } \label{sec_app:resonant_a_state}

For resonant excitation of the state $\vert a \rangle$, i.e. $\delta_1 =0$ , the population amplitude is given by
\begin{align}
c^{\text{res}}_a(t) &= - \frac{\mu_{ag}}{2 \pi i \hbar} \bigg( \pi \tilde{\mathcal{E}}(0) - i  \fint_{-\infty}^{\infty} \frac{\tilde{\mathcal{E}}(\omega) e^{i \omega t}}{ \omega} \text{d}\omega    \bigg).
\end{align}
Since the spectrum $\tilde{\mathcal{E}}(\omega)$ is symmetric, the real part of the integral vanishes and we find
\begin{align}
c^{\text{res}}_a(t) &= - \frac{\mu_{ag}}{2 \pi i \hbar} \bigg( \pi \tilde{\mathcal{E}}(0) +  \fint_{-\infty}^{\infty} \frac{\tilde{\mathcal{E}}(\omega) \sin( \omega t)}{ \omega} \text{d}\omega    \bigg).
\end{align}
For $t \rightarrow \pm \infty$ the integral is identified with $\pm \pi \tilde{\mathcal{E}}(0)$ by using the definition of the $\delta$-distribution \cite{Bracewell:2000,Goodman:1996:xi}. Hence, in the resonant case we obtain the known expressions
\begin{align}
&c^{\text{res}}_a(- \infty) =0 ~~\text{and}~~ c^{\text{res}}_a(\infty) = - \frac{\mu_{ag}}{i \hbar} \tilde{\mathcal{E}}(0).
\end{align}
In the following, we consider the non-resonant excitation of the state $\vert a \rangle $, since the equivalent $4p$-state in the experiment is also excited only near- but non-resonantly (cf. mechanical blocker in Sec.~\ref{sec:experiment}).

\subsubsection{Non-resonant excitation of the state $\vert a \rangle$: }\label{sec_app:a_state_nonres}

For the non-resonant excitation of $\vert a \rangle$, i.e. if the detuning $\delta_1$ exceeds the spectral width $\Delta \omega$, it follows $\tilde{\mathcal{E}}(\delta_1) \approx 0$ such that the population amplitude from Eq.~\eqref{eq_app:a_state_full} reads
\begin{equation}
c^{\text{non}}_a(t) \approx  \frac{\mu_{ag}}{2 \pi  \hbar}  e^{-i \delta_1 t} \fint_{-\infty}^{\infty} \frac{\tilde{\mathcal{E}}(\omega) e^{i \omega t}}{ \omega - \delta_1} \text{d}\omega .
\end{equation}
For such a large detuning $\delta_1$ we use a series expansion of the fraction in the integral $\frac{1}{\omega - \delta_1} \approx - \frac{1}{\delta_1} - \frac{\omega}{\delta^2_1} + \mathcal{O}\left( \frac{\omega^2}{\delta^3} \right)$. Thus, the principal value can be approximated by
\begin{align}
\fint_{-\infty}^{\infty} \frac{\tilde{\mathcal{E}}(\omega) e^{i \omega t}}{ \omega - \delta_1} \text{d}\omega &\approx - 2 \pi \left( \frac{1}{\delta_1} \mathcal{E}(t) - \frac{i}{\delta_1^2} \frac{\partial \mathcal{E}(t)}{\partial t} \right) \notag \\
& \approx - \frac{2\pi}{\delta_1} \mathcal{E}(t) e^{-i \zeta t},
\end{align}  
with the linear temporal phase $\zeta = \frac{\partial_t^2 \mathcal{E}(0)}{\delta_1 \mathcal{E}(0)}$, neglecting higher order phases and assuming a symmetric pulse envelope, i.e., $\partial_t \mathcal{E}(0) = 0$. For a Gaussian-shaped envelope the phase $\zeta$ takes the form $\zeta = -\frac{\Delta \omega^2}{4 \ln(2) \, \delta_1}$. Therefore the \textit{transient} population amplitude of the state $\vert a \rangle$ for non-resonant excitation is given by
\begin{equation}\label{eq_app:a_state_off}
c_a^{\text{non}}(t) \approx -\frac{\mu_{ag}}{\delta_1 \hbar} e^{-i \hat{\delta}_1 t} \mathcal{E}(t),
\end{equation} 
with the reduced detuning $\hat{\delta}_1 = \delta_1 + \zeta$. Hence, for non-resonant excitation the population amplitude $c_a(t)$ of the state $\vert a \rangle$ follows the temporal field envelope $\mathcal{E}(t)$ and decreases with the detuning via $1/\delta_1$ which also takes into account its sign.

\subsection{Excitation of the state $\vert b \rangle$}\label{sec_app:state_b_excitation}

For the excitation of the state $\vert b \rangle$ via a two-photon process we apply $2^\text{nd}$ order perturbation theory. Assuming a non-resonant excitation of state $\vert a \rangle$, we build on the results from Sec.~\ref{sec_app:state_a_excitation}, i.e., Eq.~\eqref{eq_app:a_state_off}. The population amplitude of $\vert b \rangle$ then takes the form
\begin{align}\label{eq_app:b_state_integral}
c_b(t) &= - \frac{\mu_{ba}}{i \hbar} \int_{-\infty}^{t} c^{\text{non}}_a(t') \mathcal{E}(t') e^{-i \omega_0 t'} e^{i \omega_{ba} t'} \text{d}t' \notag \\
&=\frac{\mu_{ba} \mu_{ag}}{i \delta_1 \hbar^2} \int_{-\infty}^{t} \mathcal{E}^2(t') e^{-i \delta_2 t'} \text{d}t',
\end{align}
with $\delta_2 = \hat{\delta}_1 + \omega_0 - \omega_{ba}$.
Note that the energy conservation is incorporated by the phase $e^{-i \hat{\delta}_1 t}$ from Eq.~\eqref{eq_app:a_state_off}, shifting the transition frequency in the $2^\text{nd}$ order about the detuning of the non-resonant excitation from $1^\text{st}$ order, i.e., $\omega_{ba} \rightarrow \hat{\delta}_1 + \omega_{ba}$. Following an analogous procedure as described in Sec.~\ref{sec_app:state_a_excitation} we find
\begin{align}\label{eq_app:b_state_full}
c_b(t) = \frac{\mu^{(2)}_{bg}}{2 \pi i \delta_1 \hbar^2} &\bigg( \pi \tilde{\mathcal{E}}^{(2)}(\delta_2) \notag \\
& - i   \fint_{-\infty}^{\infty} \frac{ \tilde{\mathcal{E}}^{(2)}(\omega) e^{-i ( \delta_2- \omega) t}}{ \omega - \delta_2 } \text{d}\omega  \bigg),
\end{align}
with $\mu^{(2)}_{bg} = \mu_{ba} \mu_{ag}$ as the two photon transition dipole moment.

\subsubsection{Resonant excitation of state $\vert b \rangle$: }\label{sec:app:res_b}

For two-photon resonant excitation of $\vert b \rangle$ we assume $ \delta_2 = 0$, leading to
\begin{align}
c_b^{\text{res}}(t) &= \frac{\mu^{(2)}_{bg}}{2 \pi i \delta_1 \hbar^2} \bigg( \pi \tilde{\mathcal{E}}^{(2)}(0) - i   \fint_{-\infty}^{\infty} \frac{ \tilde{\mathcal{E}}^{(2)}(\omega) e^{i \omega t}}{ \omega} \text{d}\omega \bigg) .
\end{align}
For a symmetric temporal pulse envelope $\mathcal{E}(t)$ and the MPI process to take place in in a narrow time window, we can apply a series expansion at $t=0$. In this way, we find the analytical description for the resonant two-photon probability amplitude
\begin{equation}\label{eq_app:b_state_res}
c_b^{\text{res}}(t) = \frac{\mu^{(2)}_{bg}}{2  \delta_1 \hbar^2} e^{-i \frac{\pi}{2}} \bigg(  \tilde{\mathcal{E}}^{(2)}(0) + 2 t \mathcal{E}^2(0) + \mathcal{O}(t^3) \bigg) .
\end{equation}
Note that the term in the order of $t^2$ vanishes since $\mathcal{E}(t)$ is assumed to exhibit a maximum at $t=0$. The second term in Eq.~\eqref{eq_app:b_state_res} represents the linearly time-increasing population amplitude due to laser excitation and is a sufficient approximation only for $t \approx 0$.

\subsubsection{Non-resonant excitation of state $ \vert b \rangle$: }

For non-resonant excitation of the $\vert b \rangle$ state we use $ \delta_2 \neq 0$ in Eq.~\eqref{eq_app:b_state_full}. Analogous to $1^{\text{st}}$ order perturbation theory in Sec.~\ref{sec_app:a_state_nonres} we assume a sufficiently large detuning compared to the spectral width of the second order spectrum such that $\tilde{\mathcal{E}}^{(2)}( \delta_2) \approx 0$, which leads to
\begin{equation}\label{eq_app:b_state_nonres}
c_b^{\text{non}}(t) \approx \frac{\mu_{bg}^{(2)}}{ \hbar^2  \delta_1 \delta_2} e^{-i \delta_2 t} \mathcal{E}^2(t).
\end{equation} 
Due to the large detuning $\delta_2$ in the experiment, we neglect an additional resonance induced quantum phase at this point, i.e., $\delta_2 \approx \hat{\delta}_2$. Similar to Eq.~\eqref{eq_app:a_state_off} the population amplitude $c_b^{\text{non}}(t)$ follows the quadratic temporal field envelope and decreases with the detunings via $1/(\delta_1 \delta_2)$ which also takes into account their signs.

\subsection{Excitation of the state $\vert f \rangle$ / photoionization}

To describe the photoionization of the (non\hbox{-})resonantly excited state $\vert b \rangle$ into the final free electron state $\vert f \rangle$ at energy $\varepsilon = \hbar \omega_f - \hbar \om{IP}$, we assume a perturbative one photon transition with $t \rightarrow \infty$. Since we examine both resonant and non-resonant excitation of the state $\vert b \rangle$ in the experiment (and in Sec.~\ref{sec_app:state_b_excitation}), we distinguish between these two cases also for the photoionization step.

\subsubsection{Photoionization of the resonantly excited state $\vert b \rangle$} \label{sec:app:Res_b}

To describe the photoionization of the resonantly excited state $\vert b \rangle$, we build on Eq.~\eqref{eq_app:b_state_res} which leads to
\begin{align}\label{eq:photoel}
c_f^{\infty,\text{res}}(\varepsilon) \approx  \frac{\mu_{fg}^{(3)}}{2 \hbar^3 \delta_1} \int_{-\infty}^{\infty} &\bigg( \tilde{\mathcal{E}}^{(2)}(0)  + 2 t' \mathcal{E}^2(0) \bigg) \mathcal{E}(t') e^{- i \varpi(\varepsilon) t'} \text{d}t',
\end{align}
where $\mu_{fg}^{(3)} = \mu_{fb} \mu_{bg}^{(2)} $ denotes the three-photon transition dipole moment and 
\begin{equation}
\varpi(\varepsilon) = \omega_0 - \omega_{fb}   = \omega_0 - \frac{\varepsilon}{\hbar} + \omega_b - \omega_\text{IP} 
\end{equation}
photoelectron detuning. Now, the linearly increasing population amplitude in the second term of Eq.~\eqref{eq:photoel} windows the laser electric field of the probe pulse, ensuring the assumption of $t \approx 0$ for the linear approximation in Sec.~\ref{sec:app:res_b}. This windowing results in a time-shift between the probing electric field and the probed photoelectron distribution, resulting in a linear spectral phase (cf. bottom frame in Fig.~\ref{fig:GenericScheme}(c)). Hence, the second term is equivalent to the first time-like moment of one part of the final population amplitude. Making use of the differentiation theorem for the Fourier transform \cite{Bracewell:2000,Goodman:1996:xi} we obtain the photoelectron amplitude 
\begin{equation}\label{eq:app:c_f_res}
c_f^{\infty,\text{res}}(\varepsilon) \approx \frac{\mu_{fg}^{(3)}}{ 2\hbar^3 \delta_1}  \bigg( \tilde{\mathcal{E}}^{(2)}(0)\tilde{\mathcal{E}}(\varpi(\varepsilon)) - 2i \mathcal{E}^2(0) \frac{\partial \tilde{\mathcal{E}} }{\partial \omega }(\varpi(\varepsilon)) \bigg).
\end{equation}
Around the center contribution of the photoelectron signal at $\varpi(\varepsilon)=0$, the photoelectron amplitude can be approximated by
\begin{equation}
c_f^{\infty,\text{res}}(\varepsilon) \approx \frac{\mu_{fg}^{(3)}}{ 2\hbar^3 \delta_1}   \, \tilde{\mathcal{E}}^{(2)}(0)\tilde{\mathcal{E}}(\varpi(\varepsilon)) \, e^{i \xi(\varepsilon)} ,
\end{equation}
with the additional resonance induced and energy-dependent quantum phase 
\begin{equation}
\xi(\varepsilon)  = - 2\frac{\mathcal{E}^2(0) \frac{\partial \tilde{\mathcal{E}} }{\partial \omega }(\varpi(\varepsilon))}{\tilde{\mathcal{E}}^{(2)}(0)\tilde{\mathcal{E}}(\varpi(\varepsilon))} .
\end{equation}
Using again a Gaussian-shaped envelope we find
\begin{equation}\label{eq:appen_xi}
\xi(\varepsilon) = 4 \sqrt{\frac{\ln(2)}{\pi}} \frac{\varpi(\varepsilon)}{\Delta \omega} = -\frac{\ta{res}}{\hbar} \varepsilon + \xi_0,
\end{equation}
with the resonance induced time delay $\tau_\text{res} = \sqrt{\frac{\ln(2)}{\pi}} \frac{4}{\Delta \omega} $ and an offset $\xi_0 = 4 \sqrt{\frac{\ln(2)}{\pi}} \frac{\omega_0 + \omega_b - \om{IP}}{\Delta \omega} $.

\subsubsection{Photoionization of the non-resonantly excited state $\vert b \rangle$}

Photoionization of the non-resonantly excited state $\vert b \rangle$ is based on Eq.~\eqref{eq_app:b_state_nonres} and leads to a photoelectron amplitude
\begin{align}\label{eq:app:c_f_non}
c_f^{\infty, \text{non}}(\varepsilon) &= - \frac{\mu_{fg}^{(3)}}{i \hbar^3 \delta_1  \delta_2} \int_{-\infty}^{\infty} \mathcal{E}^3(t') e^{-i \delta_2 t'} e^{-i \varpi(\varepsilon) t'} \text{d}t' \notag \\
&=i \frac{ \mu_{fg}^{(3)}}{\hbar^3 \delta_1 \delta_2}  \tilde{\mathcal{E}}^{(3)}(\varpi(\varepsilon)+  \delta_2),
\end{align}
which agrees with the results from \cite{Dudovich:2005:PRL:083002,Meshulach:1999:PRA:1287} for fully non-resonant MPI processes.\\[2.5cm]


\begin{thebibliography}{83}%
	\makeatletter
	\providecommand \@ifxundefined [1]{%
		\@ifx{#1\undefined}
	}%
	\providecommand \@ifnum [1]{%
		\ifnum #1\expandafter \@firstoftwo
		\else \expandafter \@secondoftwo
		\fi
	}%
	\providecommand \@ifx [1]{%
		\ifx #1\expandafter \@firstoftwo
		\else \expandafter \@secondoftwo
		\fi
	}%
	\providecommand \natexlab [1]{#1}%
	\providecommand \enquote  [1]{``#1''}%
	\providecommand \bibnamefont  [1]{#1}%
	\providecommand \bibfnamefont [1]{#1}%
	\providecommand \citenamefont [1]{#1}%
	\providecommand \href@noop [0]{\@secondoftwo}%
	\providecommand \href [0]{\begingroup \@sanitize@url \@href}%
	\providecommand \@href[1]{\@@startlink{#1}\@@href}%
	\providecommand \@@href[1]{\endgroup#1\@@endlink}%
	\providecommand \@sanitize@url [0]{\catcode `\\12\catcode `\$12\catcode
		`\&12\catcode `\#12\catcode `\^12\catcode `\_12\catcode `\%12\relax}%
	\providecommand \@@startlink[1]{}%
	\providecommand \@@endlink[0]{}%
	\providecommand \url  [0]{\begingroup\@sanitize@url \@url }%
	\providecommand \@url [1]{\endgroup\@href {#1}{\urlprefix }}%
	\providecommand \urlprefix  [0]{URL }%
	\providecommand \Eprint [0]{\href }%
	\providecommand \doibase [0]{https://doi.org/}%
	\providecommand \selectlanguage [0]{\@gobble}%
	\providecommand \bibinfo  [0]{\@secondoftwo}%
	\providecommand \bibfield  [0]{\@secondoftwo}%
	\providecommand \translation [1]{[#1]}%
	\providecommand \BibitemOpen [0]{}%
	\providecommand \bibitemStop [0]{}%
	\providecommand \bibitemNoStop [0]{.\EOS\space}%
	\providecommand \EOS [0]{\spacefactor3000\relax}%
	\providecommand \BibitemShut  [1]{\csname bibitem#1\endcsname}%
	\let\auto@bib@innerbib\@empty
	\bibitem [{\citenamefont {Einstein}(1905)}]{Einstein:1905:AP:165}%
	\BibitemOpen
	\bibfield  {author} {\bibinfo {author} {\bibfnamefont {A.}~\bibnamefont
			{Einstein}},\ }\bibfield  {title} {\bibinfo {title} {{\"{U}}ber einen die
			{E}rzeugung und {V}erwandlung des {L}ichtes betreffenden heuristischen
			{G}esichtspunkt},\ }\href@noop {} {\bibfield  {journal} {\bibinfo  {journal}
			{Ann. Phys.}\ }\textbf {\bibinfo {volume} {14}},\ \bibinfo {pages} {165}
		(\bibinfo {year} {1905})}\BibitemShut {NoStop}%
	\bibitem [{\citenamefont
		{G\"{o}ppert-Mayer}(1931)}]{Goeppert-Mayer:1931:AP:273}%
	\BibitemOpen
	\bibfield  {author} {\bibinfo {author} {\bibfnamefont {M.}~\bibnamefont
			{G\"{o}ppert-Mayer}},\ }\bibfield  {title} {\bibinfo {title} {\"{U}ber
			{E}lementarakte mit zwei {Q}uantenspr\"{u}ngen},\ }\href@noop {} {\bibfield
		{journal} {\bibinfo  {journal} {Ann. Phys.}\ }\textbf {\bibinfo {volume}
			{401}},\ \bibinfo {pages} {273} (\bibinfo {year} {1931})}\BibitemShut
	{NoStop}%
	\bibitem [{\citenamefont {Lambropoulos}(1976)}]{Lambropoulos:1976:87}%
	\BibitemOpen
	\bibfield  {author} {\bibinfo {author} {\bibfnamefont {P.}~\bibnamefont
			{Lambropoulos}},\ }\bibinfo {title} {Topics on multiphoton processes in
		atoms},\ in\ \href@noop {} {\emph {\bibinfo {booktitle} {Advances in atomic
				and molecular physics}}}\ (\bibinfo {year} {1976})\ pp.\ \bibinfo {pages} {87
		--164}\BibitemShut {NoStop}%
	\bibitem [{\citenamefont {Faisal}(1987)}]{Faisal:1987}%
	\BibitemOpen
	\bibfield  {author} {\bibinfo {author} {\bibfnamefont {F.~H.~M.}\
			\bibnamefont {Faisal}},\ }\href@noop {} {\emph {\bibinfo {title} {Theory of
				multiphoton processes}}}\ (\bibinfo  {publisher} {Springer Science \&
		Business Media},\ \bibinfo {year} {1987})\BibitemShut {NoStop}%
	\bibitem [{\citenamefont {Fujimura}\ and\ \citenamefont
		{Lin}(2003)}]{Fujimura:2003:199}%
	\BibitemOpen
	\bibfield  {author} {\bibinfo {author} {\bibfnamefont {Y.}~\bibnamefont
			{Fujimura}}\ and\ \bibinfo {author} {\bibfnamefont {S.~H.}\ \bibnamefont
			{Lin}},\ }\bibinfo {title} {Multiphoton spectroscopy},\ in\ \href@noop {}
	{\emph {\bibinfo {booktitle} {Encyclopedia of Physical Science and Technology
				(Third Edition)}}},\ \bibinfo {editor} {edited by\ \bibinfo {editor}
		{\bibfnamefont {R.~A.}\ \bibnamefont {Meyers}}}\ (\bibinfo  {publisher}
	{Academic Press},\ \bibinfo {address} {New York},\ \bibinfo {year} {2003})\
	pp.\ \bibinfo {pages} {199--229}\BibitemShut {NoStop}%
	\bibitem [{\citenamefont {Letokhov}(1987)}]{Letokhov:1987}%
	\BibitemOpen
	\bibfield  {author} {\bibinfo {author} {\bibfnamefont {V.}~\bibnamefont
			{Letokhov}},\ }\href@noop {} {\emph {\bibinfo {title} {Laser photoionization
				spectroscopy}}}\ (\bibinfo  {publisher} {Elsevier},\ \bibinfo {year}
	{1987})\BibitemShut {NoStop}%
	\bibitem [{\citenamefont {Kitzler}\ and\ \citenamefont
		{Gr\"{a}fe}(2016)}]{Kitzler:2016}%
	\BibitemOpen
	\bibfield  {author} {\bibinfo {author} {\bibfnamefont {M.}~\bibnamefont
			{Kitzler}}\ and\ \bibinfo {author} {\bibfnamefont {S.}~\bibnamefont
			{Gr\"{a}fe}},\ }\href@noop {} {\emph {\bibinfo {title} {Ultrafast Dynamics
				Driven by Intense Light Pulses}}},\ \bibinfo {series} {Springer Series on
		Atomic, Optical, and Plasma Physics}, Vol.~\bibinfo {volume} {86}\ (\bibinfo
	{year} {2016})\BibitemShut {NoStop}%
	\bibitem [{\citenamefont {Lin}\ \emph {et~al.}(2016)\citenamefont {Lin},
		\citenamefont {Villaeys},\ and\ \citenamefont {Fujimura}}]{Lin:2016}%
	\BibitemOpen
	\bibfield  {author} {\bibinfo {author} {\bibfnamefont {S.~H.}\ \bibnamefont
			{Lin}}, \bibinfo {author} {\bibfnamefont {A.~A.}\ \bibnamefont {Villaeys}},\
		and\ \bibinfo {author} {\bibfnamefont {Y.}~\bibnamefont {Fujimura}},\
	}\href@noop {} {\emph {\bibinfo {title} {Advances in Multi-Photon Processes
			and Spectroscopy}}},\ \bibinfo {series} {Advances in Multi-Photon Processes
	and Spectroscopy}, Vol.~\bibinfo {volume} {23}\ (\bibinfo  {publisher} {World
	Scientific},\ \bibinfo {address} {Singapore, Hackensack, London},\ \bibinfo
{year} {2016})\BibitemShut {NoStop}%
\bibitem [{\citenamefont {Bauer}(2017)}]{Bauer:2017}%
\BibitemOpen
\bibfield  {author} {\bibinfo {author} {\bibfnamefont {D.}~\bibnamefont
		{Bauer}},\ }\href@noop {} {\emph {\bibinfo {title} {Computational
			strong-field quantum dynamics. Intense light-matter interactions}}},\
Vol.~\bibinfo {volume} {1}\ (\bibinfo  {publisher} {De Gruyter},\ \bibinfo
{year} {2017})\BibitemShut {NoStop}%
\bibitem [{\citenamefont {Hockett}(2018)}]{Hockett:2018a}%
\BibitemOpen
\bibfield  {author} {\bibinfo {author} {\bibfnamefont {P.}~\bibnamefont
		{Hockett}},\ }\href@noop {} {\emph {\bibinfo {title} {Quantum Metrology with
			Photoelectrons: Volume I: Foundations}}}\ (\bibinfo  {publisher} {Morgan \&
	Claypool Publishers},\ \bibinfo {year} {2018})\BibitemShut {NoStop}%
\bibitem [{\citenamefont {Yamanouchi}\ \emph {et~al.}(2021)\citenamefont
	{Yamanouchi}, \citenamefont {Midorikawa},\ and\ \citenamefont
	{Roso}}]{Yamanouchi:2021}%
\BibitemOpen
\bibfield  {author} {\bibinfo {author} {\bibfnamefont {K.}~\bibnamefont
		{Yamanouchi}}, \bibinfo {author} {\bibfnamefont {K.}~\bibnamefont
		{Midorikawa}},\ and\ \bibinfo {author} {\bibfnamefont {L.}~\bibnamefont
		{Roso}},\ }\href@noop {} {\emph {\bibinfo {title} {Progress in Ultrafast
			Intense Laser Science XVI}}}\ (\bibinfo  {publisher} {Springer},\ \bibinfo
{year} {2021})\BibitemShut {NoStop}%
\bibitem [{\citenamefont {Fedorov}\ and\ \citenamefont
	{Kazakov}(1989)}]{Fedorov:1989:PQE:1}%
\BibitemOpen
\bibfield  {author} {\bibinfo {author} {\bibfnamefont {M.~V.}\ \bibnamefont
		{Fedorov}}\ and\ \bibinfo {author} {\bibfnamefont {A.~E.}\ \bibnamefont
		{Kazakov}},\ }\bibfield  {title} {\bibinfo {title} {Resonances and saturation
		in multiphoton bound-free transitions},\ }\href@noop {} {\bibfield  {journal}
	{\bibinfo  {journal} {Prog. Quant. Electr.}\ }\textbf {\bibinfo {volume}
		{13}},\ \bibinfo {pages} {1} (\bibinfo {year} {1989})}\BibitemShut {NoStop}%
\bibitem [{\citenamefont {Mainfray}\ and\ \citenamefont
	{Manus}(1991)}]{Mainfray:1991:RPP:1333}%
\BibitemOpen
\bibfield  {author} {\bibinfo {author} {\bibfnamefont {G.}~\bibnamefont
		{Mainfray}}\ and\ \bibinfo {author} {\bibfnamefont {C.}~\bibnamefont
		{Manus}},\ }\bibfield  {title} {\bibinfo {title} {Multiphoton ionization of
		atoms},\ }\href@noop {} {\bibfield  {journal} {\bibinfo  {journal} {Rep.
			Prog. Phys.}\ }\textbf {\bibinfo {volume} {54}},\ \bibinfo {pages} {1333}
	(\bibinfo {year} {1991})}\BibitemShut {NoStop}%
\bibitem [{\citenamefont {D\"{o}rner}\ \emph {et~al.}(2000)\citenamefont
	{D\"{o}rner}, \citenamefont {Mergel}, \citenamefont {Jagutzki}, \citenamefont
	{Spielberger}, \citenamefont {Ullrich}, \citenamefont {Moshammer},\ and\
	\citenamefont {Schmidt-Bocking}}]{Doerner:2000:PR:95}%
\BibitemOpen
\bibfield  {author} {\bibinfo {author} {\bibfnamefont {R.}~\bibnamefont
		{D\"{o}rner}}, \bibinfo {author} {\bibfnamefont {V.}~\bibnamefont {Mergel}},
	\bibinfo {author} {\bibfnamefont {O.}~\bibnamefont {Jagutzki}}, \bibinfo
	{author} {\bibfnamefont {L.}~\bibnamefont {Spielberger}}, \bibinfo {author}
	{\bibfnamefont {J.}~\bibnamefont {Ullrich}}, \bibinfo {author} {\bibfnamefont
		{R.}~\bibnamefont {Moshammer}},\ and\ \bibinfo {author} {\bibfnamefont
		{H.}~\bibnamefont {Schmidt-Bocking}},\ }\bibfield  {title} {\bibinfo {title}
	{Cold target recoil ion momentum spectroscopy: a 'momentum microscope' to
		view atomic collision dynamics},\ }\href@noop {} {\bibfield  {journal}
	{\bibinfo  {journal} {Phys. Rep.}\ }\textbf {\bibinfo {volume} {330}},\
	\bibinfo {pages} {95} (\bibinfo {year} {2000})}\BibitemShut {NoStop}%
\bibitem [{\citenamefont {Wollenhaupt}\ \emph {et~al.}(2005)\citenamefont
	{Wollenhaupt}, \citenamefont {Engel},\ and\ \citenamefont
	{Baumert}}]{Wollenhaupt:2005:ARPC:25}%
\BibitemOpen
\bibfield  {author} {\bibinfo {author} {\bibfnamefont {M.}~\bibnamefont
		{Wollenhaupt}}, \bibinfo {author} {\bibfnamefont {V.}~\bibnamefont {Engel}},\
	and\ \bibinfo {author} {\bibfnamefont {T.}~\bibnamefont {Baumert}},\
}\bibfield  {title} {\bibinfo {title} {Femtosecond laser photoelectron
	spectroscopy on atoms and small molecules: Prototype studies in quantum
	control},\ }\href@noop {} {\bibfield  {journal} {\bibinfo  {journal} {Annu.
		Rev. Phys. Chem.}\ }\textbf {\bibinfo {volume} {56}},\ \bibinfo {pages} {25}
(\bibinfo {year} {2005})}\BibitemShut {NoStop}%
\bibitem [{\citenamefont {Reid}(2012)}]{Reid:2012:MP:131}%
\BibitemOpen
\bibfield  {author} {\bibinfo {author} {\bibfnamefont {K.~L.}\ \bibnamefont
		{Reid}},\ }\bibfield  {title} {\bibinfo {title} {Photoelectron angular
		distributions: developments in applications to isolated molecular systems},\
}\href@noop {} {\bibfield  {journal} {\bibinfo  {journal} {Mol.Phys.}\
}\textbf {\bibinfo {volume} {110}},\ \bibinfo {pages} {131} (\bibinfo {year}
{2012})}\BibitemShut {NoStop}%
\bibitem [{\citenamefont {Goulielmakis}\ \emph {et~al.}(2010)\citenamefont
	{Goulielmakis}, \citenamefont {Loh}, \citenamefont {Wirth}, \citenamefont
	{Santra}, \citenamefont {Rohringer}, \citenamefont {Yakovlev}, \citenamefont
	{Zherebtsov}, \citenamefont {Pfeifer}, \citenamefont {Azzeer}, \citenamefont
	{Kling}, \citenamefont {Leone},\ and\ \citenamefont
	{Krausz}}]{Goulielmakis:2010:Nature:739}%
\BibitemOpen
\bibfield  {author} {\bibinfo {author} {\bibfnamefont {E.}~\bibnamefont
		{Goulielmakis}}, \bibinfo {author} {\bibfnamefont {Z.~H.}\ \bibnamefont
		{Loh}}, \bibinfo {author} {\bibfnamefont {A.}~\bibnamefont {Wirth}}, \bibinfo
	{author} {\bibfnamefont {R.}~\bibnamefont {Santra}}, \bibinfo {author}
	{\bibfnamefont {N.}~\bibnamefont {Rohringer}}, \bibinfo {author}
	{\bibfnamefont {V.~S.}\ \bibnamefont {Yakovlev}}, \bibinfo {author}
	{\bibfnamefont {S.}~\bibnamefont {Zherebtsov}}, \bibinfo {author}
	{\bibfnamefont {T.}~\bibnamefont {Pfeifer}}, \bibinfo {author} {\bibfnamefont
		{A.~M.}\ \bibnamefont {Azzeer}}, \bibinfo {author} {\bibfnamefont {M.~F.}\
		\bibnamefont {Kling}}, \bibinfo {author} {\bibfnamefont {S.~R.}\ \bibnamefont
		{Leone}},\ and\ \bibinfo {author} {\bibfnamefont {F.}~\bibnamefont
		{Krausz}},\ }\bibfield  {title} {\bibinfo {title} {Real-time observation of
		valence electron motion},\ }\href@noop {} {\bibfield  {journal} {\bibinfo
		{journal} {Nature}\ }\textbf {\bibinfo {volume} {466}},\ \bibinfo {pages}
	{739} (\bibinfo {year} {2010})}\BibitemShut {NoStop}%
\bibitem [{\citenamefont {Villeneuve}\ \emph {et~al.}(2017)\citenamefont
	{Villeneuve}, \citenamefont {Hockett}, \citenamefont {Vrakking},\ and\
	\citenamefont {Niikura}}]{Villeneuve:2017:Science:1150}%
\BibitemOpen
\bibfield  {author} {\bibinfo {author} {\bibfnamefont {D.~M.}\ \bibnamefont
		{Villeneuve}}, \bibinfo {author} {\bibfnamefont {P.}~\bibnamefont {Hockett}},
	\bibinfo {author} {\bibfnamefont {M.~J.~J.}\ \bibnamefont {Vrakking}},\ and\
	\bibinfo {author} {\bibfnamefont {H.}~\bibnamefont {Niikura}},\ }\bibfield
{title} {\bibinfo {title} {Coherent imaging of an attosecond electron wave
		packet},\ }\href@noop {} {\bibfield  {journal} {\bibinfo  {journal}
		{Science}\ }\textbf {\bibinfo {volume} {356}},\ \bibinfo {pages} {1150}
	(\bibinfo {year} {2017})}\BibitemShut {NoStop}%
\bibitem [{\citenamefont {Rice}\ and\ \citenamefont
	{Zhao}(2000)}]{Rice:2000:456}%
\BibitemOpen
\bibfield  {author} {\bibinfo {author} {\bibfnamefont {S.~A.}\ \bibnamefont
		{Rice}}\ and\ \bibinfo {author} {\bibfnamefont {M.}~\bibnamefont {Zhao}},\
}\href@noop {} {\emph {\bibinfo {title} {Optical Control of Molecular
		Dynamics}}}\ (\bibinfo  {publisher} {Wiley},\ \bibinfo {address} {New York},\
\bibinfo {year} {2000})\ p.\ \bibinfo {pages} {456}\BibitemShut {NoStop}%
\bibitem [{\citenamefont {Goswami}(2003)}]{Goswami:2003:PR:385}%
\BibitemOpen
\bibfield  {author} {\bibinfo {author} {\bibfnamefont {D.}~\bibnamefont
		{Goswami}},\ }\bibfield  {title} {\bibinfo {title} {Optical pulse shaping
		approaches to coherent control},\ }\href@noop {} {\bibfield  {journal}
	{\bibinfo  {journal} {Phys. Rep.}\ }\textbf {\bibinfo {volume} {374}},\
	\bibinfo {pages} {385} (\bibinfo {year} {2003})}\BibitemShut {NoStop}%
\bibitem [{\citenamefont {Brif}\ \emph {et~al.}(2010)\citenamefont {Brif},
	\citenamefont {Chakrabarti},\ and\ \citenamefont
	{Rabitz}}]{Brif:2010:NJP:075008}%
\BibitemOpen
\bibfield  {author} {\bibinfo {author} {\bibfnamefont {C.}~\bibnamefont
		{Brif}}, \bibinfo {author} {\bibfnamefont {R.}~\bibnamefont {Chakrabarti}},\
	and\ \bibinfo {author} {\bibfnamefont {H.}~\bibnamefont {Rabitz}},\
}\bibfield  {title} {\bibinfo {title} {Control of quantum phenomena: past,
	present and future},\ }\href@noop {} {\bibfield  {journal} {\bibinfo
	{journal} {New J. Phys.}\ }\textbf {\bibinfo {volume} {12}},\ \bibinfo
{pages} {075008} (\bibinfo {year} {2010})}\BibitemShut {NoStop}%
\bibitem [{\citenamefont {Shapiro}\ and\ \citenamefont
	{Brumer}(2012)}]{Shapiro:2012:562}%
\BibitemOpen
\bibfield  {author} {\bibinfo {author} {\bibfnamefont {M.}~\bibnamefont
		{Shapiro}}\ and\ \bibinfo {author} {\bibfnamefont {P.}~\bibnamefont
		{Brumer}},\ }\href@noop {} {\emph {\bibinfo {title} {Quantum Control of
			Molecular Processes}}},\ \bibinfo {edition} {2nd}\ ed.\ (\bibinfo
{publisher} {Wiley},\ \bibinfo {address} {New York},\ \bibinfo {year}
{2012})\ p.\ \bibinfo {pages} {562}\BibitemShut {NoStop}%
\bibitem [{\citenamefont {Wollenhaupt}\ \emph {et~al.}(2015)\citenamefont
	{Wollenhaupt}, \citenamefont {Bayer},\ and\ \citenamefont
	{Baumert}}]{Wollenhaupt:2015:63}%
\BibitemOpen
\bibfield  {author} {\bibinfo {author} {\bibfnamefont {M.}~\bibnamefont
		{Wollenhaupt}}, \bibinfo {author} {\bibfnamefont {T.}~\bibnamefont {Bayer}},\
	and\ \bibinfo {author} {\bibfnamefont {T.}~\bibnamefont {Baumert}},\
}\bibinfo {title} {Control of ultrafast electron dynamics with shaped
femtosecond laser pulses: From atoms to solids},\ in\ \href@noop {} {\emph
{\bibinfo {booktitle} {Ultrafast Dynamics Driven by Intense Light Pulses}}},\
\bibinfo {editor} {edited by\ \bibinfo {editor} {\bibfnamefont
		{S.}~\bibnamefont {Gr\"{a}fe}}\ and\ \bibinfo {editor} {\bibfnamefont
		{M.}~\bibnamefont {Kitzler}}}\ (\bibinfo  {publisher} {Springer},\ \bibinfo
{year} {2015})\ \bibinfo {type} {Book section}~\bibinfo {chapter} {4}, pp.\
\bibinfo {pages} {63--122}\BibitemShut {NoStop}%
\bibitem [{\citenamefont {Trallero-Herrero}\ \emph {et~al.}(2006)\citenamefont
	{Trallero-Herrero}, \citenamefont {Cohen},\ and\ \citenamefont
	{Weinacht}}]{Trallero-Herrero:2006:PRL:063603}%
\BibitemOpen
\bibfield  {author} {\bibinfo {author} {\bibfnamefont {C.}~\bibnamefont
		{Trallero-Herrero}}, \bibinfo {author} {\bibfnamefont {J.~L.}\ \bibnamefont
		{Cohen}},\ and\ \bibinfo {author} {\bibfnamefont {T.}~\bibnamefont
		{Weinacht}},\ }\bibfield  {title} {\bibinfo {title} {Strong-field atomic
		phase matching},\ }\href@noop {} {\bibfield  {journal} {\bibinfo  {journal}
		{Phys. Rev. Lett.}\ }\textbf {\bibinfo {volume} {96}},\ \bibinfo {pages}
	{063603 } (\bibinfo {year} {2006})}\BibitemShut {NoStop}%
\bibitem [{\citenamefont {Ngoko~Djiokap}\ \emph {et~al.}(2015)\citenamefont
	{Ngoko~Djiokap}, \citenamefont {Hu}, \citenamefont {Madsen}, \citenamefont
	{Manakov}, \citenamefont {Meremianin},\ and\ \citenamefont
	{Starace}}]{NgokoDjiokap:2015:PRL:113004}%
\BibitemOpen
\bibfield  {author} {\bibinfo {author} {\bibfnamefont {J.~M.}\ \bibnamefont
		{Ngoko~Djiokap}}, \bibinfo {author} {\bibfnamefont {S.~X.}\ \bibnamefont
		{Hu}}, \bibinfo {author} {\bibfnamefont {L.~B.}\ \bibnamefont {Madsen}},
	\bibinfo {author} {\bibfnamefont {N.~L.}\ \bibnamefont {Manakov}}, \bibinfo
	{author} {\bibfnamefont {A.~V.}\ \bibnamefont {Meremianin}},\ and\ \bibinfo
	{author} {\bibfnamefont {A.~F.}\ \bibnamefont {Starace}},\ }\bibfield
{title} {\bibinfo {title} {Electron vortices in photoionization by circularly
		polarized attosecond pulses},\ }\href@noop {} {\bibfield  {journal} {\bibinfo
		{journal} {Phys. Rev. Lett.}\ }\textbf {\bibinfo {volume} {115}},\ \bibinfo
	{pages} {113004} (\bibinfo {year} {2015})}\BibitemShut {NoStop}%
\bibitem [{\citenamefont {Yuan}\ \emph {et~al.}(2016)\citenamefont {Yuan},
	\citenamefont {Chelkowski},\ and\ \citenamefont
	{Bandrauk}}]{Yuan:2016:PRA:053425}%
\BibitemOpen
\bibfield  {author} {\bibinfo {author} {\bibfnamefont {K.-J.}\ \bibnamefont
		{Yuan}}, \bibinfo {author} {\bibfnamefont {S.}~\bibnamefont {Chelkowski}},\
	and\ \bibinfo {author} {\bibfnamefont {A.~D.}\ \bibnamefont {Bandrauk}},\
}\bibfield  {title} {\bibinfo {title} {Photoelectron momentum distributions
	of molecules in bichromatic circularly polarized attosecond {U}{V} laser
	fields},\ }\href@noop {} {\bibfield  {journal} {\bibinfo  {journal} {Phys.
		Rev. A}\ }\textbf {\bibinfo {volume} {93}},\ \bibinfo {pages} {053425}
(\bibinfo {year} {2016})}\BibitemShut {NoStop}%
\bibitem [{\citenamefont {Pengel}\ \emph
	{et~al.}(2017{\natexlab{a}})\citenamefont {Pengel}, \citenamefont
	{Kerbstadt}, \citenamefont {Johannmeyer}, \citenamefont {Englert},
	\citenamefont {Bayer},\ and\ \citenamefont
	{Wollenhaupt}}]{Pengel:2017:PRL:053003}%
\BibitemOpen
\bibfield  {author} {\bibinfo {author} {\bibfnamefont {D.}~\bibnamefont
		{Pengel}}, \bibinfo {author} {\bibfnamefont {S.}~\bibnamefont {Kerbstadt}},
	\bibinfo {author} {\bibfnamefont {D.}~\bibnamefont {Johannmeyer}}, \bibinfo
	{author} {\bibfnamefont {L.}~\bibnamefont {Englert}}, \bibinfo {author}
	{\bibfnamefont {T.}~\bibnamefont {Bayer}},\ and\ \bibinfo {author}
	{\bibfnamefont {M.}~\bibnamefont {Wollenhaupt}},\ }\bibfield  {title}
{\bibinfo {title} {Electron vortices in femtosecond multiphoton ionization},\
}\href@noop {} {\bibfield  {journal} {\bibinfo  {journal} {Phys. Rev. Lett.}\
}\textbf {\bibinfo {volume} {118}},\ \bibinfo {pages} {053003} (\bibinfo
{year} {2017}{\natexlab{a}})}\BibitemShut {NoStop}%
\bibitem [{\citenamefont {Kerbstadt}\ \emph
	{et~al.}(2019{\natexlab{a}})\citenamefont {Kerbstadt}, \citenamefont
	{Eickhoff}, \citenamefont {Bayer},\ and\ \citenamefont
	{Wollenhaupt}}]{Kerbstadt:2019:NC:658}%
\BibitemOpen
\bibfield  {author} {\bibinfo {author} {\bibfnamefont {S.}~\bibnamefont
		{Kerbstadt}}, \bibinfo {author} {\bibfnamefont {K.}~\bibnamefont {Eickhoff}},
	\bibinfo {author} {\bibfnamefont {T.}~\bibnamefont {Bayer}},\ and\ \bibinfo
	{author} {\bibfnamefont {M.}~\bibnamefont {Wollenhaupt}},\ }\bibfield
{title} {\bibinfo {title} {Odd electron wave packets from cycloidal
		ultrashort laser fields},\ }\href@noop {} {\bibfield  {journal} {\bibinfo
		{journal} {Nat. Comm.}\ }\textbf {\bibinfo {volume} {10}},\ \bibinfo {pages}
	{658} (\bibinfo {year} {2019}{\natexlab{a}})}\BibitemShut {NoStop}%
\bibitem [{\citenamefont {Eickhoff}\ \emph
	{et~al.}(2021{\natexlab{a}})\citenamefont {Eickhoff}, \citenamefont
	{Englert}, \citenamefont {Bayer},\ and\ \citenamefont
	{Wollenhaupt}}]{Eickhoff:2021:FP:444}%
\BibitemOpen
\bibfield  {author} {\bibinfo {author} {\bibfnamefont {K.}~\bibnamefont
		{Eickhoff}}, \bibinfo {author} {\bibfnamefont {L.}~\bibnamefont {Englert}},
	\bibinfo {author} {\bibfnamefont {T.}~\bibnamefont {Bayer}},\ and\ \bibinfo
	{author} {\bibfnamefont {M.}~\bibnamefont {Wollenhaupt}},\ }\bibfield
{title} {\bibinfo {title} {Multichromatic polarization-controlled pulse
		sequences for coherent control of multiphoton ionization},\ }\href@noop {}
{\bibfield  {journal} {\bibinfo  {journal} {Front. Phys.}\ }\textbf {\bibinfo
		{volume} {9}},\ \bibinfo {pages} {444} (\bibinfo {year}
	{2021}{\natexlab{a}})}\BibitemShut {NoStop}%
\bibitem [{\citenamefont {Wollenhaupt}\ \emph {et~al.}(2002)\citenamefont
	{Wollenhaupt}, \citenamefont {Assion}, \citenamefont {Liese}, \citenamefont
	{Sarpe-Tudoran}, \citenamefont {Baumert}, \citenamefont {Zamith},
	\citenamefont {Bouchene}, \citenamefont {Girard}, \citenamefont {Flettner},
	\citenamefont {Weichmann},\ and\ \citenamefont
	{Gerber}}]{Wollenhaupt:2002:PRL:173001}%
\BibitemOpen
\bibfield  {author} {\bibinfo {author} {\bibfnamefont {M.}~\bibnamefont
		{Wollenhaupt}}, \bibinfo {author} {\bibfnamefont {A.}~\bibnamefont {Assion}},
	\bibinfo {author} {\bibfnamefont {D.}~\bibnamefont {Liese}}, \bibinfo
	{author} {\bibfnamefont {C.}~\bibnamefont {Sarpe-Tudoran}}, \bibinfo {author}
	{\bibfnamefont {T.}~\bibnamefont {Baumert}}, \bibinfo {author} {\bibfnamefont
		{S.}~\bibnamefont {Zamith}}, \bibinfo {author} {\bibfnamefont {M.~A.}\
		\bibnamefont {Bouchene}}, \bibinfo {author} {\bibfnamefont {B.}~\bibnamefont
		{Girard}}, \bibinfo {author} {\bibfnamefont {A.}~\bibnamefont {Flettner}},
	\bibinfo {author} {\bibfnamefont {U.}~\bibnamefont {Weichmann}},\ and\
	\bibinfo {author} {\bibfnamefont {G.}~\bibnamefont {Gerber}},\ }\bibfield
{title} {\bibinfo {title} {Interferences of ultrashort free electron wave
		packets},\ }\href@noop {} {\bibfield  {journal} {\bibinfo  {journal} {Phys.
			Rev. Lett.}\ }\textbf {\bibinfo {volume} {89}},\ \bibinfo {pages} {173001}
	(\bibinfo {year} {2002})}\BibitemShut {NoStop}%
\bibitem [{\citenamefont {Warren}\ \emph {et~al.}(1993)\citenamefont {Warren},
	\citenamefont {Rabitz},\ and\ \citenamefont
	{Dahleh}}]{Warren:1993:Science:1581}%
\BibitemOpen
\bibfield  {author} {\bibinfo {author} {\bibfnamefont {W.~S.}\ \bibnamefont
		{Warren}}, \bibinfo {author} {\bibfnamefont {H.}~\bibnamefont {Rabitz}},\
	and\ \bibinfo {author} {\bibfnamefont {M.}~\bibnamefont {Dahleh}},\
}\bibfield  {title} {\bibinfo {title} {Coherent control of quantum dynamics:
	The dream is alive},\ }\href@noop {} {\bibfield  {journal} {\bibinfo
	{journal} {Science}\ }\textbf {\bibinfo {volume} {259}},\ \bibinfo {pages}
{1581} (\bibinfo {year} {1993})}\BibitemShut {NoStop}%
\bibitem [{\citenamefont {Rabitz}\ \emph {et~al.}(2000)\citenamefont {Rabitz},
	\citenamefont {de~Vivie-Riedle}, \citenamefont {Motzkus},\ and\ \citenamefont
	{Kompa}}]{Rabitz:2000:Science:824}%
\BibitemOpen
\bibfield  {author} {\bibinfo {author} {\bibfnamefont {H.}~\bibnamefont
		{Rabitz}}, \bibinfo {author} {\bibfnamefont {R.}~\bibnamefont
		{de~Vivie-Riedle}}, \bibinfo {author} {\bibfnamefont {M.}~\bibnamefont
		{Motzkus}},\ and\ \bibinfo {author} {\bibfnamefont {K.}~\bibnamefont
		{Kompa}},\ }\bibfield  {title} {\bibinfo {title} {Whither the future of
		controlling quantum phenomena?},\ }\href@noop {} {\bibfield  {journal}
	{\bibinfo  {journal} {Science}\ }\textbf {\bibinfo {volume} {288}},\ \bibinfo
	{pages} {824} (\bibinfo {year} {2000})}\BibitemShut {NoStop}%
\bibitem [{\citenamefont {Rabitz}(2011)}]{Rabitz:2011:FD:415}%
\BibitemOpen
\bibfield  {author} {\bibinfo {author} {\bibfnamefont {H.}~\bibnamefont
		{Rabitz}},\ }\bibfield  {title} {\bibinfo {title} {A perspective on
		controlling quantum phenomena},\ }\href@noop {} {\bibfield  {journal}
	{\bibinfo  {journal} {Faraday Discuss.}\ }\textbf {\bibinfo {volume} {153}},\
	\bibinfo {pages} {415} (\bibinfo {year} {2011})}\BibitemShut {NoStop}%
\bibitem [{\citenamefont {Weinacht}\ \emph {et~al.}(1999)\citenamefont
	{Weinacht}, \citenamefont {Ahn},\ and\ \citenamefont
	{Bucksbaum}}]{Weinacht:1999:Nature:233}%
\BibitemOpen
\bibfield  {author} {\bibinfo {author} {\bibfnamefont {T.~C.}\ \bibnamefont
		{Weinacht}}, \bibinfo {author} {\bibfnamefont {J.}~\bibnamefont {Ahn}},\ and\
	\bibinfo {author} {\bibfnamefont {P.~H.}\ \bibnamefont {Bucksbaum}},\
}\bibfield  {title} {\bibinfo {title} {Controlling the shape of a quantum
	wavefunction},\ }\href@noop {} {\bibfield  {journal} {\bibinfo  {journal}
	{Nature}\ }\textbf {\bibinfo {volume} {397}},\ \bibinfo {pages} {233}
(\bibinfo {year} {1999})}\BibitemShut {NoStop}%
\bibitem [{\citenamefont {Kaufman}\ \emph {et~al.}(2020)\citenamefont
	{Kaufman}, \citenamefont {Rozgonyi}, \citenamefont {Marquetand},\ and\
	\citenamefont {Weinacht}}]{Kaufman:2020:PRL:053202}%
\BibitemOpen
\bibfield  {author} {\bibinfo {author} {\bibfnamefont {B.}~\bibnamefont
		{Kaufman}}, \bibinfo {author} {\bibfnamefont {T.}~\bibnamefont {Rozgonyi}},
	\bibinfo {author} {\bibfnamefont {P.}~\bibnamefont {Marquetand}},\ and\
	\bibinfo {author} {\bibfnamefont {T.}~\bibnamefont {Weinacht}},\ }\bibfield
{title} {\bibinfo {title} {Coherent control of internal conversion in
		strong-field molecular ionization},\ }\href@noop {} {\bibfield  {journal}
	{\bibinfo  {journal} {Phys. Rev. Lett.}\ }\textbf {\bibinfo {volume} {125}},\
	\bibinfo {pages} {053202} (\bibinfo {year} {2020})}\BibitemShut {NoStop}%
\bibitem [{\citenamefont {Leichtle}\ \emph {et~al.}(1998)\citenamefont
	{Leichtle}, \citenamefont {Schleich}, \citenamefont {Averbukh},\ and\
	\citenamefont {Shapiro}}]{Leichtle:1998:PRL:1418}%
\BibitemOpen
\bibfield  {author} {\bibinfo {author} {\bibfnamefont {C.}~\bibnamefont
		{Leichtle}}, \bibinfo {author} {\bibfnamefont {W.~P.}\ \bibnamefont
		{Schleich}}, \bibinfo {author} {\bibfnamefont {I.~S.}\ \bibnamefont
		{Averbukh}},\ and\ \bibinfo {author} {\bibfnamefont {M.}~\bibnamefont
		{Shapiro}},\ }\bibfield  {title} {\bibinfo {title} {Quantum state
		holography},\ }\href@noop {} {\bibfield  {journal} {\bibinfo  {journal}
		{Phys. Rev. Lett.}\ }\textbf {\bibinfo {volume} {80}},\ \bibinfo {pages}
	{1418} (\bibinfo {year} {1998})}\BibitemShut {NoStop}%
\bibitem [{\citenamefont {Averbukh}\ \emph {et~al.}(1999)\citenamefont
	{Averbukh}, \citenamefont {Shapiro}, \citenamefont {Leichtle},\ and\
	\citenamefont {Schleich}}]{Averbukh:1999:PRA:2163}%
\BibitemOpen
\bibfield  {author} {\bibinfo {author} {\bibfnamefont {I.~S.}\ \bibnamefont
		{Averbukh}}, \bibinfo {author} {\bibfnamefont {M.}~\bibnamefont {Shapiro}},
	\bibinfo {author} {\bibfnamefont {C.}~\bibnamefont {Leichtle}},\ and\
	\bibinfo {author} {\bibfnamefont {W.~P.}\ \bibnamefont {Schleich}},\
}\bibfield  {title} {\bibinfo {title} {Reconstructing wave packets by
	quantum-state holography},\ }\href@noop {} {\bibfield  {journal} {\bibinfo
	{journal} {Phys. Rev. A}\ }\textbf {\bibinfo {volume} {59}},\ \bibinfo
{pages} {2163} (\bibinfo {year} {1999})}\BibitemShut {NoStop}%
\bibitem [{\citenamefont {de~Morisson~Faria}\ and\ \citenamefont
	{Maxwell}(2020)}]{deMorissonFaria:2020:RPP:034401}%
\BibitemOpen
\bibfield  {author} {\bibinfo {author} {\bibfnamefont {C.~F.}\ \bibnamefont
		{de~Morisson~Faria}}\ and\ \bibinfo {author} {\bibfnamefont {A.~S.}\
		\bibnamefont {Maxwell}},\ }\bibfield  {title} {\bibinfo {title} {It is all
		about phases: ultrafast holographic photoelectron imaging},\ }\href@noop {}
{\bibfield  {journal} {\bibinfo  {journal} {Rep. Prog. Phys.}\ }\textbf
	{\bibinfo {volume} {83}},\ \bibinfo {pages} {034401} (\bibinfo {year}
	{2020})}\BibitemShut {NoStop}%
\bibitem [{\citenamefont {Isinger}\ \emph {et~al.}(2017)\citenamefont
	{Isinger}, \citenamefont {Squibb}, \citenamefont {Busto}, \citenamefont
	{Zhong}, \citenamefont {Harth}, \citenamefont {Kroon}, \citenamefont {Nandi},
	\citenamefont {Arnold}, \citenamefont {Miranda},\ and\ \citenamefont
	{Dahlstr\"{o}m}}]{Isinger:2017:Science:893}%
\BibitemOpen
\bibfield  {author} {\bibinfo {author} {\bibfnamefont {M.}~\bibnamefont
		{Isinger}}, \bibinfo {author} {\bibfnamefont {R.~J.}\ \bibnamefont {Squibb}},
	\bibinfo {author} {\bibfnamefont {D.}~\bibnamefont {Busto}}, \bibinfo
	{author} {\bibfnamefont {S.}~\bibnamefont {Zhong}}, \bibinfo {author}
	{\bibfnamefont {A.}~\bibnamefont {Harth}}, \bibinfo {author} {\bibfnamefont
		{D.}~\bibnamefont {Kroon}}, \bibinfo {author} {\bibfnamefont
		{S.}~\bibnamefont {Nandi}}, \bibinfo {author} {\bibfnamefont {C.~L.}\
		\bibnamefont {Arnold}}, \bibinfo {author} {\bibfnamefont {M.}~\bibnamefont
		{Miranda}},\ and\ \bibinfo {author} {\bibfnamefont {J.~M.}\ \bibnamefont
		{Dahlstr\"{o}m}},\ }\bibfield  {title} {\bibinfo {title} {Photoionization in
		the time and frequency domain},\ }\href@noop {} {\bibfield  {journal}
	{\bibinfo  {journal} {Science}\ }\textbf {\bibinfo {volume} {358}},\ \bibinfo
	{pages} {893} (\bibinfo {year} {2017})}\BibitemShut {NoStop}%
\bibitem [{\citenamefont {Swoboda}\ \emph {et~al.}(2010)\citenamefont
	{Swoboda}, \citenamefont {Fordell}, \citenamefont {Kl\"{u}nder},
	\citenamefont {Dahlstr\"{o}m}, \citenamefont {Miranda}, \citenamefont {Buth},
	\citenamefont {Schafer}, \citenamefont {Mauritsson}, \citenamefont
	{L'Huillier},\ and\ \citenamefont {Gisselbrecht}}]{Swoboda:2010:PRL:103003}%
\BibitemOpen
\bibfield  {author} {\bibinfo {author} {\bibfnamefont {M.}~\bibnamefont
		{Swoboda}}, \bibinfo {author} {\bibfnamefont {T.}~\bibnamefont {Fordell}},
	\bibinfo {author} {\bibfnamefont {K.}~\bibnamefont {Kl\"{u}nder}}, \bibinfo
	{author} {\bibfnamefont {J.~M.}\ \bibnamefont {Dahlstr\"{o}m}}, \bibinfo
	{author} {\bibfnamefont {M.}~\bibnamefont {Miranda}}, \bibinfo {author}
	{\bibfnamefont {C.}~\bibnamefont {Buth}}, \bibinfo {author} {\bibfnamefont
		{K.~J.}\ \bibnamefont {Schafer}}, \bibinfo {author} {\bibfnamefont
		{J.}~\bibnamefont {Mauritsson}}, \bibinfo {author} {\bibfnamefont
		{A.}~\bibnamefont {L'Huillier}},\ and\ \bibinfo {author} {\bibfnamefont
		{M.}~\bibnamefont {Gisselbrecht}},\ }\bibfield  {title} {\bibinfo {title}
	{Phase measurement of resonant two-photon ionization in helium},\ }\href@noop
{} {\bibfield  {journal} {\bibinfo  {journal} {Phys. Rev. Lett.}\ }\textbf
	{\bibinfo {volume} {104}},\ \bibinfo {pages} {103003} (\bibinfo {year}
	{2010})}\BibitemShut {NoStop}%
\bibitem [{\citenamefont {Drescher}\ \emph {et~al.}(2022)\citenamefont
	{Drescher}, \citenamefont {Witting}, \citenamefont {Kornilov},\ and\
	\citenamefont {Vrakking}}]{Drescher:2022:PRA:L011101}%
\BibitemOpen
\bibfield  {author} {\bibinfo {author} {\bibfnamefont {L.}~\bibnamefont
		{Drescher}}, \bibinfo {author} {\bibfnamefont {T.}~\bibnamefont {Witting}},
	\bibinfo {author} {\bibfnamefont {O.}~\bibnamefont {Kornilov}},\ and\
	\bibinfo {author} {\bibfnamefont {M.~J.~J.}\ \bibnamefont {Vrakking}},\
}\bibfield  {title} {\bibinfo {title} {Phase dependence of resonant and
	antiresonant two-photon excitations},\ }\href@noop {} {\bibfield  {journal}
{\bibinfo  {journal} {Phys. Rev. A}\ }\textbf {\bibinfo {volume} {105}},\
\bibinfo {pages} {L011101} (\bibinfo {year} {2022})}\BibitemShut {NoStop}%
\bibitem [{\citenamefont {Haessler}\ \emph {et~al.}(2009)\citenamefont
	{Haessler}, \citenamefont {Fabre}, \citenamefont {Higuet}, \citenamefont
	{Caillat}, \citenamefont {Ruchon}, \citenamefont {Breger}, \citenamefont
	{Carr\'{e}}, \citenamefont {Constant}, \citenamefont {Maquet},\ and\
	\citenamefont {M\'{e}vel}}]{Haessler:2009:PRA:011404}%
\BibitemOpen
\bibfield  {author} {\bibinfo {author} {\bibfnamefont {S.}~\bibnamefont
		{Haessler}}, \bibinfo {author} {\bibfnamefont {B.}~\bibnamefont {Fabre}},
	\bibinfo {author} {\bibfnamefont {J.}~\bibnamefont {Higuet}}, \bibinfo
	{author} {\bibfnamefont {J.}~\bibnamefont {Caillat}}, \bibinfo {author}
	{\bibfnamefont {T.}~\bibnamefont {Ruchon}}, \bibinfo {author} {\bibfnamefont
		{P.}~\bibnamefont {Breger}}, \bibinfo {author} {\bibfnamefont
		{B.}~\bibnamefont {Carr\'{e}}}, \bibinfo {author} {\bibfnamefont
		{E.}~\bibnamefont {Constant}}, \bibinfo {author} {\bibfnamefont
		{A.}~\bibnamefont {Maquet}},\ and\ \bibinfo {author} {\bibfnamefont
		{E.}~\bibnamefont {M\'{e}vel}},\ }\bibfield  {title} {\bibinfo {title}
	{Phase-resolved attosecond near-threshold photoionization of molecular
		nitrogen},\ }\href@noop {} {\bibfield  {journal} {\bibinfo  {journal} {Phys.
			Rev. A}\ }\textbf {\bibinfo {volume} {80}},\ \bibinfo {pages} {011404}
	(\bibinfo {year} {2009})}\BibitemShut {NoStop}%
\bibitem [{\citenamefont {Vos}\ \emph {et~al.}(2018)\citenamefont {Vos},
	\citenamefont {Cattaneo}, \citenamefont {Patchkovskii}, \citenamefont
	{Zimmermann}, \citenamefont {Cirelli}, \citenamefont {Lucchini},
	\citenamefont {Kheifets}, \citenamefont {Landsman},\ and\ \citenamefont
	{Keller}}]{Vos:2018:Science:1326}%
\BibitemOpen
\bibfield  {author} {\bibinfo {author} {\bibfnamefont {J.}~\bibnamefont
		{Vos}}, \bibinfo {author} {\bibfnamefont {L.}~\bibnamefont {Cattaneo}},
	\bibinfo {author} {\bibfnamefont {S.}~\bibnamefont {Patchkovskii}}, \bibinfo
	{author} {\bibfnamefont {T.}~\bibnamefont {Zimmermann}}, \bibinfo {author}
	{\bibfnamefont {C.}~\bibnamefont {Cirelli}}, \bibinfo {author} {\bibfnamefont
		{M.}~\bibnamefont {Lucchini}}, \bibinfo {author} {\bibfnamefont
		{A.}~\bibnamefont {Kheifets}}, \bibinfo {author} {\bibfnamefont {A.~S.}\
		\bibnamefont {Landsman}},\ and\ \bibinfo {author} {\bibfnamefont
		{U.}~\bibnamefont {Keller}},\ }\bibfield  {title} {\bibinfo {title}
	{Orientation-dependent stereo wigner time delay and electron localization in
		a small molecule},\ }\href@noop {} {\bibfield  {journal} {\bibinfo  {journal}
		{Science}\ }\textbf {\bibinfo {volume} {360}},\ \bibinfo {pages} {1326}
	(\bibinfo {year} {2018})}\BibitemShut {NoStop}%
\bibitem [{\citenamefont {Grundmann}\ \emph {et~al.}(2020)\citenamefont
	{Grundmann}, \citenamefont {Trabert}, \citenamefont {Fehre}, \citenamefont
	{Strenger}, \citenamefont {Pier}, \citenamefont {Kaiser}, \citenamefont
	{Kircher}, \citenamefont {Weller}, \citenamefont {Eckart},\ and\
	\citenamefont {Schmidt}}]{Grundmann:2020:Science:339}%
\BibitemOpen
\bibfield  {author} {\bibinfo {author} {\bibfnamefont {S.}~\bibnamefont
		{Grundmann}}, \bibinfo {author} {\bibfnamefont {D.}~\bibnamefont {Trabert}},
	\bibinfo {author} {\bibfnamefont {K.}~\bibnamefont {Fehre}}, \bibinfo
	{author} {\bibfnamefont {N.}~\bibnamefont {Strenger}}, \bibinfo {author}
	{\bibfnamefont {A.}~\bibnamefont {Pier}}, \bibinfo {author} {\bibfnamefont
		{L.}~\bibnamefont {Kaiser}}, \bibinfo {author} {\bibfnamefont
		{M.}~\bibnamefont {Kircher}}, \bibinfo {author} {\bibfnamefont
		{M.}~\bibnamefont {Weller}}, \bibinfo {author} {\bibfnamefont
		{S.}~\bibnamefont {Eckart}},\ and\ \bibinfo {author} {\bibfnamefont
		{L.~P.~H.}\ \bibnamefont {Schmidt}},\ }\bibfield  {title} {\bibinfo {title}
	{Zeptosecond birth time delay in molecular photoionization},\ }\href@noop {}
{\bibfield  {journal} {\bibinfo  {journal} {Science}\ }\textbf {\bibinfo
		{volume} {370}},\ \bibinfo {pages} {339} (\bibinfo {year}
	{2020})}\BibitemShut {NoStop}%
\bibitem [{\citenamefont {Gruson}\ \emph {et~al.}(2016)\citenamefont {Gruson},
	\citenamefont {Barreau}, \citenamefont {Jim\'{e}nez-Galan}, \citenamefont
	{Risoud}, \citenamefont {Caillat}, \citenamefont {Maquet}, \citenamefont
	{Carr\'{e}}, \citenamefont {Lepetit}, \citenamefont {Hergott}, \citenamefont
	{Ruchon}, \citenamefont {Argenti}, \citenamefont {Ta\"{i}eb}, \citenamefont
	{Mart\'{i}n},\ and\ \citenamefont {Sali\`{e}res}}]{Gruson:2016:Science:734}%
\BibitemOpen
\bibfield  {author} {\bibinfo {author} {\bibfnamefont {V.}~\bibnamefont
		{Gruson}}, \bibinfo {author} {\bibfnamefont {L.}~\bibnamefont {Barreau}},
	\bibinfo {author} {\bibfnamefont {A.}~\bibnamefont {Jim\'{e}nez-Galan}},
	\bibinfo {author} {\bibfnamefont {F.}~\bibnamefont {Risoud}}, \bibinfo
	{author} {\bibfnamefont {J.}~\bibnamefont {Caillat}}, \bibinfo {author}
	{\bibfnamefont {A.}~\bibnamefont {Maquet}}, \bibinfo {author} {\bibfnamefont
		{B.}~\bibnamefont {Carr\'{e}}}, \bibinfo {author} {\bibfnamefont
		{F.}~\bibnamefont {Lepetit}}, \bibinfo {author} {\bibfnamefont {J.~F.}\
		\bibnamefont {Hergott}}, \bibinfo {author} {\bibfnamefont {T.}~\bibnamefont
		{Ruchon}}, \bibinfo {author} {\bibfnamefont {L.}~\bibnamefont {Argenti}},
	\bibinfo {author} {\bibfnamefont {R.}~\bibnamefont {Ta\"{i}eb}}, \bibinfo
	{author} {\bibfnamefont {F.}~\bibnamefont {Mart\'{i}n}},\ and\ \bibinfo
	{author} {\bibfnamefont {P.}~\bibnamefont {Sali\`{e}res}},\ }\bibfield
{title} {\bibinfo {title} {Attosecond dynamics through a fano resonance:
		Monitoring the birth of a photoelectron},\ }\href@noop {} {\bibfield
	{journal} {\bibinfo  {journal} {Science}\ }\textbf {\bibinfo {volume}
		{354}},\ \bibinfo {pages} {734} (\bibinfo {year} {2016})}\BibitemShut
{NoStop}%
\bibitem [{\citenamefont {Hofmann}\ \emph {et~al.}(2019)\citenamefont
	{Hofmann}, \citenamefont {Landsman},\ and\ \citenamefont
	{Keller}}]{Hofmann:2019:JMO:1052}%
\BibitemOpen
\bibfield  {author} {\bibinfo {author} {\bibfnamefont {C.}~\bibnamefont
		{Hofmann}}, \bibinfo {author} {\bibfnamefont {A.~S.}\ \bibnamefont
		{Landsman}},\ and\ \bibinfo {author} {\bibfnamefont {U.}~\bibnamefont
		{Keller}},\ }\bibfield  {title} {\bibinfo {title} {Attoclock revisited on
		electron tunnelling time},\ }\href@noop {} {\bibfield  {journal} {\bibinfo
		{journal} {Journal of Modern Optics}\ }\textbf {\bibinfo {volume} {66}},\
	\bibinfo {pages} {1052} (\bibinfo {year} {2019})}\BibitemShut {NoStop}%
\bibitem [{\citenamefont {Sokolovski}\ and\ \citenamefont
	{Akhmatskaya}(2018)}]{Sokolovski:2018:CP:1}%
\BibitemOpen
\bibfield  {author} {\bibinfo {author} {\bibfnamefont {D.}~\bibnamefont
		{Sokolovski}}\ and\ \bibinfo {author} {\bibfnamefont {E.}~\bibnamefont
		{Akhmatskaya}},\ }\bibfield  {title} {\bibinfo {title} {No time at the end of
		the tunnel},\ }\href@noop {} {\bibfield  {journal} {\bibinfo  {journal}
		{Communications Physics}\ }\textbf {\bibinfo {volume} {1}},\ \bibinfo {pages}
	{1} (\bibinfo {year} {2018})}\BibitemShut {NoStop}%
\bibitem [{\citenamefont {Hockett}\ \emph {et~al.}(2016)\citenamefont
	{Hockett}, \citenamefont {Frumker}, \citenamefont {Villeneuve},\ and\
	\citenamefont {Corkum}}]{Hockett:2016:JPBAMOP:095602}%
\BibitemOpen
\bibfield  {author} {\bibinfo {author} {\bibfnamefont {P.}~\bibnamefont
		{Hockett}}, \bibinfo {author} {\bibfnamefont {E.}~\bibnamefont {Frumker}},
	\bibinfo {author} {\bibfnamefont {D.~M.}\ \bibnamefont {Villeneuve}},\ and\
	\bibinfo {author} {\bibfnamefont {P.~B.}\ \bibnamefont {Corkum}},\ }\bibfield
{title} {\bibinfo {title} {Time delay in molecular photoionization},\
}\href@noop {} {\bibfield  {journal} {\bibinfo  {journal} {J. Phys. B: At.
		Mol. Opt. Phys.}\ }\textbf {\bibinfo {volume} {49}},\ \bibinfo {pages}
{095602} (\bibinfo {year} {2016})}\BibitemShut {NoStop}%
\bibitem [{\citenamefont {Harth}\ \emph {et~al.}(2019)\citenamefont {Harth},
	\citenamefont {Douguet}, \citenamefont {Bartschat}, \citenamefont
	{Moshammer},\ and\ \citenamefont {Pfeifer}}]{Harth:2019:PRA:023410}%
\BibitemOpen
\bibfield  {author} {\bibinfo {author} {\bibfnamefont {A.}~\bibnamefont
		{Harth}}, \bibinfo {author} {\bibfnamefont {N.}~\bibnamefont {Douguet}},
	\bibinfo {author} {\bibfnamefont {K.}~\bibnamefont {Bartschat}}, \bibinfo
	{author} {\bibfnamefont {R.}~\bibnamefont {Moshammer}},\ and\ \bibinfo
	{author} {\bibfnamefont {T.}~\bibnamefont {Pfeifer}},\ }\bibfield  {title}
{\bibinfo {title} {Extracting phase information on continuum-continuum
		couplings},\ }\href@noop {} {\bibfield  {journal} {\bibinfo  {journal} {Phys.
			Rev. A}\ }\textbf {\bibinfo {volume} {99}},\ \bibinfo {pages} {023410}
	(\bibinfo {year} {2019})}\BibitemShut {NoStop}%
\bibitem [{\citenamefont {Dahlstr\"{o}m}\ \emph {et~al.}(2013)\citenamefont
	{Dahlstr\"{o}m}, \citenamefont {Gu\'{e}not}, \citenamefont {Kl\"{u}nder},
	\citenamefont {Gisselbrecht}, \citenamefont {Mauritsson}, \citenamefont
	{L'Huillier}, \citenamefont {Maquet},\ and\ \citenamefont
	{Ta\"{i}eb}}]{Dahlstroem:2013:CP:53}%
\BibitemOpen
\bibfield  {author} {\bibinfo {author} {\bibfnamefont {J.~M.}\ \bibnamefont
		{Dahlstr\"{o}m}}, \bibinfo {author} {\bibfnamefont {D.}~\bibnamefont
		{Gu\'{e}not}}, \bibinfo {author} {\bibfnamefont {K.}~\bibnamefont
		{Kl\"{u}nder}}, \bibinfo {author} {\bibfnamefont {M.}~\bibnamefont
		{Gisselbrecht}}, \bibinfo {author} {\bibfnamefont {J.}~\bibnamefont
		{Mauritsson}}, \bibinfo {author} {\bibfnamefont {A.}~\bibnamefont
		{L'Huillier}}, \bibinfo {author} {\bibfnamefont {A.}~\bibnamefont {Maquet}},\
	and\ \bibinfo {author} {\bibfnamefont {R.}~\bibnamefont {Ta\"{i}eb}},\
}\bibfield  {title} {\bibinfo {title} {Theory of attosecond delays in
	laser-assisted photoionization},\ }\href@noop {} {\bibfield  {journal}
{\bibinfo  {journal} {Chem. Phys.}\ }\textbf {\bibinfo {volume} {414}},\
\bibinfo {pages} {53} (\bibinfo {year} {2013})}\BibitemShut {NoStop}%
\bibitem [{\citenamefont {Schultze}\ \emph {et~al.}(2010)\citenamefont
	{Schultze}, \citenamefont {Fie\'{a}}, \citenamefont {Karpowicz},
	\citenamefont {Gagnon}, \citenamefont {Korbman}, \citenamefont {Hofstetter},
	\citenamefont {Neppl}, \citenamefont {Cavalieri}, \citenamefont {Komninos},
	\citenamefont {Mercouris}, \citenamefont {Nocolaides}, \citenamefont
	{Pazourek}, \citenamefont {Nagele}, \citenamefont {Feist}, \citenamefont
	{Burgd\"{o}rfer}, \citenamefont {Azzeer}, \citenamefont {Ernstofer},
	\citenamefont {Kienberger}, \citenamefont {Kleineberg}, \citenamefont
	{Goulielmakis}, \citenamefont {Krausz},\ and\ \citenamefont
	{Yakovlev}}]{Schultze:2010:Science:1685}%
\BibitemOpen
\bibfield  {author} {\bibinfo {author} {\bibfnamefont {M.}~\bibnamefont
		{Schultze}}, \bibinfo {author} {\bibfnamefont {M.}~\bibnamefont {Fie\'{a}}},
	\bibinfo {author} {\bibfnamefont {N.}~\bibnamefont {Karpowicz}}, \bibinfo
	{author} {\bibfnamefont {J.}~\bibnamefont {Gagnon}}, \bibinfo {author}
	{\bibfnamefont {M.}~\bibnamefont {Korbman}}, \bibinfo {author} {\bibfnamefont
		{M.}~\bibnamefont {Hofstetter}}, \bibinfo {author} {\bibfnamefont
		{S.}~\bibnamefont {Neppl}}, \bibinfo {author} {\bibfnamefont {A.~L.}\
		\bibnamefont {Cavalieri}}, \bibinfo {author} {\bibfnamefont {Y.}~\bibnamefont
		{Komninos}}, \bibinfo {author} {\bibfnamefont {T.}~\bibnamefont {Mercouris}},
	\bibinfo {author} {\bibfnamefont {C.~A.}\ \bibnamefont {Nocolaides}},
	\bibinfo {author} {\bibfnamefont {R.}~\bibnamefont {Pazourek}}, \bibinfo
	{author} {\bibfnamefont {S.}~\bibnamefont {Nagele}}, \bibinfo {author}
	{\bibfnamefont {J.}~\bibnamefont {Feist}}, \bibinfo {author} {\bibfnamefont
		{J.}~\bibnamefont {Burgd\"{o}rfer}}, \bibinfo {author} {\bibfnamefont
		{A.~M.}\ \bibnamefont {Azzeer}}, \bibinfo {author} {\bibfnamefont
		{R.}~\bibnamefont {Ernstofer}}, \bibinfo {author} {\bibfnamefont
		{R.}~\bibnamefont {Kienberger}}, \bibinfo {author} {\bibfnamefont
		{U.}~\bibnamefont {Kleineberg}}, \bibinfo {author} {\bibfnamefont
		{E.}~\bibnamefont {Goulielmakis}}, \bibinfo {author} {\bibfnamefont
		{F.}~\bibnamefont {Krausz}},\ and\ \bibinfo {author} {\bibfnamefont {V.~S.}\
		\bibnamefont {Yakovlev}},\ }\bibfield  {title} {\bibinfo {title} {Delay in
		photoemission},\ }\href@noop {} {\bibfield  {journal} {\bibinfo  {journal}
		{Science}\ }\textbf {\bibinfo {volume} {328}},\ \bibinfo {pages} {1685}
	(\bibinfo {year} {2010})}\BibitemShut {NoStop}%
\bibitem [{\citenamefont {Dahlstr\"{o}m}\ \emph {et~al.}(2012)\citenamefont
	{Dahlstr\"{o}m}, \citenamefont {L'Huillier},\ and\ \citenamefont
	{Maquet}}]{Dahlstroem:2012:JPB:183001}%
\BibitemOpen
\bibfield  {author} {\bibinfo {author} {\bibfnamefont {J.~M.}\ \bibnamefont
		{Dahlstr\"{o}m}}, \bibinfo {author} {\bibfnamefont {A.}~\bibnamefont
		{L'Huillier}},\ and\ \bibinfo {author} {\bibfnamefont {A.}~\bibnamefont
		{Maquet}},\ }\bibfield  {title} {\bibinfo {title} {Introduction to attosecond
		delays in photoionization},\ }\href@noop {} {\bibfield  {journal} {\bibinfo
		{journal} {J. Phys. B}\ }\textbf {\bibinfo {volume} {45}},\ \bibinfo {pages}
	{183001} (\bibinfo {year} {2012})}\BibitemShut {NoStop}%
\bibitem [{\citenamefont {Eickhoff}\ \emph
	{et~al.}(2021{\natexlab{b}})\citenamefont {Eickhoff}, \citenamefont {Feld},
	\citenamefont {K\"{o}hnke}, \citenamefont {Bayer},\ and\ \citenamefont
	{Wollenhaupt}}]{Eickhoff:2021:PRA:052805}%
\BibitemOpen
\bibfield  {author} {\bibinfo {author} {\bibfnamefont {K.}~\bibnamefont
		{Eickhoff}}, \bibinfo {author} {\bibfnamefont {L.}~\bibnamefont {Feld}},
	\bibinfo {author} {\bibfnamefont {D.}~\bibnamefont {K\"{o}hnke}}, \bibinfo
	{author} {\bibfnamefont {T.}~\bibnamefont {Bayer}},\ and\ \bibinfo {author}
	{\bibfnamefont {M.}~\bibnamefont {Wollenhaupt}},\ }\bibfield  {title}
{\bibinfo {title} {Trichromatic shaper-based quantum state holography},\
}\href@noop {} {\bibfield  {journal} {\bibinfo  {journal} {Phys. Rev. A}\
}\textbf {\bibinfo {volume} {104}},\ \bibinfo {pages} {052805} (\bibinfo
{year} {2021}{\natexlab{b}})}\BibitemShut {NoStop}%
\bibitem [{\citenamefont {Kerbstadt}\ \emph
	{et~al.}(2017{\natexlab{a}})\citenamefont {Kerbstadt}, \citenamefont
	{Pengel}, \citenamefont {Johannmeyer}, \citenamefont {Englert}, \citenamefont
	{Bayer},\ and\ \citenamefont {Wollenhaupt}}]{Kerbstadt:2017:NJP:103017}%
\BibitemOpen
\bibfield  {author} {\bibinfo {author} {\bibfnamefont {S.}~\bibnamefont
		{Kerbstadt}}, \bibinfo {author} {\bibfnamefont {D.}~\bibnamefont {Pengel}},
	\bibinfo {author} {\bibfnamefont {D.}~\bibnamefont {Johannmeyer}}, \bibinfo
	{author} {\bibfnamefont {L.}~\bibnamefont {Englert}}, \bibinfo {author}
	{\bibfnamefont {T.}~\bibnamefont {Bayer}},\ and\ \bibinfo {author}
	{\bibfnamefont {M.}~\bibnamefont {Wollenhaupt}},\ }\bibfield  {title}
{\bibinfo {title} {Control of photoelectron momentum distributions by
		bichromatic polarization-shaped laser fields},\ }\href@noop {} {\bibfield
	{journal} {\bibinfo  {journal} {New J. Phys.}\ }\textbf {\bibinfo {volume}
		{19}},\ \bibinfo {pages} {103017} (\bibinfo {year}
	{2017}{\natexlab{a}})}\BibitemShut {NoStop}%
\bibitem [{\citenamefont {Kramida}\ \emph {et~al.}(2018)\citenamefont
	{Kramida}, \citenamefont {Rachenko},\ and\ \citenamefont
	{Reader}}]{Kramida:2018}%
\BibitemOpen
\bibfield  {author} {\bibinfo {author} {\bibfnamefont {A.}~\bibnamefont
		{Kramida}}, \bibinfo {author} {\bibfnamefont {Y.}~\bibnamefont {Rachenko}},\
	and\ \bibinfo {author} {\bibfnamefont {J.}~\bibnamefont {Reader}},\
}\bibfield  {title} {\bibinfo {title} {Nist atomic spectra database},\
}\href@noop {} {\bibfield  {journal} {\bibinfo  {journal} {National Institute
		of Standards and Technology}\ } (\bibinfo {year} {2018})}\BibitemShut
{NoStop}%
\bibitem [{\citenamefont {Wollenhaupt}\ \emph {et~al.}(2003)\citenamefont
	{Wollenhaupt}, \citenamefont {Assion}, \citenamefont {Bazhan}, \citenamefont
	{Horn}, \citenamefont {Liese}, \citenamefont {Sarpe-Tudoran}, \citenamefont
	{Winter},\ and\ \citenamefont {Baumert}}]{Wollenhaupt:2003:PRA:015401}%
\BibitemOpen
\bibfield  {author} {\bibinfo {author} {\bibfnamefont {M.}~\bibnamefont
		{Wollenhaupt}}, \bibinfo {author} {\bibfnamefont {A.}~\bibnamefont {Assion}},
	\bibinfo {author} {\bibfnamefont {O.}~\bibnamefont {Bazhan}}, \bibinfo
	{author} {\bibfnamefont {C.}~\bibnamefont {Horn}}, \bibinfo {author}
	{\bibfnamefont {D.}~\bibnamefont {Liese}}, \bibinfo {author} {\bibfnamefont
		{C.}~\bibnamefont {Sarpe-Tudoran}}, \bibinfo {author} {\bibfnamefont
		{M.}~\bibnamefont {Winter}},\ and\ \bibinfo {author} {\bibfnamefont
		{T.}~\bibnamefont {Baumert}},\ }\bibfield  {title} {\bibinfo {title} {Control
		of interferences in an autler-townes doublet: Symmetry of control
		parameters},\ }\href@noop {} {\bibfield  {journal} {\bibinfo  {journal}
		{Phys. Rev. A}\ }\textbf {\bibinfo {volume} {68}},\ \bibinfo {pages} {015401}
	(\bibinfo {year} {2003})}\BibitemShut {NoStop}%
\bibitem [{\citenamefont {Garcia}\ \emph {et~al.}(2004)\citenamefont {Garcia},
	\citenamefont {Nahon},\ and\ \citenamefont {Powis}}]{Garcia:2004:RSI:4989}%
\BibitemOpen
\bibfield  {author} {\bibinfo {author} {\bibfnamefont {G.~A.}\ \bibnamefont
		{Garcia}}, \bibinfo {author} {\bibfnamefont {L.}~\bibnamefont {Nahon}},\ and\
	\bibinfo {author} {\bibfnamefont {I.}~\bibnamefont {Powis}},\ }\bibfield
{title} {\bibinfo {title} {Two-dimensional charged particle image inversion
		using a polar basis function expansion},\ }\href@noop {} {\bibfield
	{journal} {\bibinfo  {journal} {Rev. Sci. Instrum.}\ }\textbf {\bibinfo
		{volume} {75}},\ \bibinfo {pages} {4989} (\bibinfo {year}
	{2004})}\BibitemShut {NoStop}%
\bibitem [{\citenamefont {Diels}\ and\ \citenamefont
	{Rudolph}(1996)}]{Diels:1996}%
\BibitemOpen
\bibfield  {author} {\bibinfo {author} {\bibfnamefont {J.~C.}\ \bibnamefont
		{Diels}}\ and\ \bibinfo {author} {\bibfnamefont {W.}~\bibnamefont
		{Rudolph}},\ }\href@noop {} {\emph {\bibinfo {title} {Ultrashort Laser Pulse
			Phenomena}}}\ (\bibinfo  {publisher} {Academic Press},\ \bibinfo {address}
{San Diego},\ \bibinfo {year} {1996})\BibitemShut {NoStop}%
\bibitem [{\citenamefont {Eickhoff}\ \emph {et~al.}(2020)\citenamefont
	{Eickhoff}, \citenamefont {Kerbstadt}, \citenamefont {Bayer},\ and\
	\citenamefont {Wollenhaupt}}]{Eickhoff:2020:PRA:013430}%
\BibitemOpen
\bibfield  {author} {\bibinfo {author} {\bibfnamefont {K.}~\bibnamefont
		{Eickhoff}}, \bibinfo {author} {\bibfnamefont {S.}~\bibnamefont {Kerbstadt}},
	\bibinfo {author} {\bibfnamefont {T.}~\bibnamefont {Bayer}},\ and\ \bibinfo
	{author} {\bibfnamefont {M.}~\bibnamefont {Wollenhaupt}},\ }\bibfield
{title} {\bibinfo {title} {Dynamic quantum state holography},\ }\href@noop {}
{\bibfield  {journal} {\bibinfo  {journal} {Phys. Rev. A}\ }\textbf {\bibinfo
		{volume} {101}},\ \bibinfo {pages} {013430} (\bibinfo {year}
	{2020})}\BibitemShut {NoStop}%
\bibitem [{\citenamefont {Bayer}\ \emph {et~al.}(2016)\citenamefont {Bayer},
	\citenamefont {Wollenhaupt}, \citenamefont {Braun},\ and\ \citenamefont
	{Baumert}}]{Bayer:2016:ACP:235}%
\BibitemOpen
\bibfield  {author} {\bibinfo {author} {\bibfnamefont {T.}~\bibnamefont
		{Bayer}}, \bibinfo {author} {\bibfnamefont {M.}~\bibnamefont {Wollenhaupt}},
	\bibinfo {author} {\bibfnamefont {H.}~\bibnamefont {Braun}},\ and\ \bibinfo
	{author} {\bibfnamefont {T.}~\bibnamefont {Baumert}},\ }\bibfield  {title}
{\bibinfo {title} {Ultrafast and efficient control of coherent electron
		dynamics via spods},\ }\href@noop {} {\bibfield  {journal} {\bibinfo
		{journal} {Adv. Chem. Phys.}\ }\textbf {\bibinfo {volume} {159}},\ \bibinfo
	{pages} {235} (\bibinfo {year} {2016})}\BibitemShut {NoStop}%
\bibitem [{\citenamefont {Kerbstadt}\ \emph
	{et~al.}(2017{\natexlab{b}})\citenamefont {Kerbstadt}, \citenamefont
	{Englert}, \citenamefont {Bayer},\ and\ \citenamefont
	{Wollenhaupt}}]{Kerbstadt:2017:JMO:1010}%
\BibitemOpen
\bibfield  {author} {\bibinfo {author} {\bibfnamefont {S.}~\bibnamefont
		{Kerbstadt}}, \bibinfo {author} {\bibfnamefont {L.}~\bibnamefont {Englert}},
	\bibinfo {author} {\bibfnamefont {T.}~\bibnamefont {Bayer}},\ and\ \bibinfo
	{author} {\bibfnamefont {M.}~\bibnamefont {Wollenhaupt}},\ }\bibfield
{title} {\bibinfo {title} {Ultrashort polarization-tailored bichromatic
		fields},\ }\href@noop {} {\bibfield  {journal} {\bibinfo  {journal} {J. Mod.
			Opt.}\ }\textbf {\bibinfo {volume} {64}},\ \bibinfo {pages} {1010} (\bibinfo
	{year} {2017}{\natexlab{b}})}\BibitemShut {NoStop}%
\bibitem [{\citenamefont {Kerbstadt}\ \emph
	{et~al.}(2017{\natexlab{c}})\citenamefont {Kerbstadt}, \citenamefont
	{Timmer}, \citenamefont {Englert}, \citenamefont {Bayer},\ and\ \citenamefont
	{Wollenhaupt}}]{Kerbstadt:2017:OE:12518}%
\BibitemOpen
\bibfield  {author} {\bibinfo {author} {\bibfnamefont {S.}~\bibnamefont
		{Kerbstadt}}, \bibinfo {author} {\bibfnamefont {D.}~\bibnamefont {Timmer}},
	\bibinfo {author} {\bibfnamefont {L.}~\bibnamefont {Englert}}, \bibinfo
	{author} {\bibfnamefont {T.}~\bibnamefont {Bayer}},\ and\ \bibinfo {author}
	{\bibfnamefont {M.}~\bibnamefont {Wollenhaupt}},\ }\bibfield  {title}
{\bibinfo {title} {Ultrashort polarization-tailored bichromatic fields from a
		{C}{E}{P}-stable white light supercontinuum},\ }\href@noop {} {\bibfield
	{journal} {\bibinfo  {journal} {Opt. Express}\ }\textbf {\bibinfo {volume}
		{25}},\ \bibinfo {pages} {12518} (\bibinfo {year}
	{2017}{\natexlab{c}})}\BibitemShut {NoStop}%
\bibitem [{\citenamefont {Eppink}\ and\ \citenamefont
	{Parker}(1997)}]{Eppink:1997:RSI:3477}%
\BibitemOpen
\bibfield  {author} {\bibinfo {author} {\bibfnamefont {A.~T. J.~B.}\
		\bibnamefont {Eppink}}\ and\ \bibinfo {author} {\bibfnamefont {D.~H.}\
		\bibnamefont {Parker}},\ }\bibfield  {title} {\bibinfo {title} {Velocity map
		imaging of ions and electrons using electrostatic lenses: Application in
		photoelectron and photofragment ion imaging of molecular oxygen},\
}\href@noop {} {\bibfield  {journal} {\bibinfo  {journal} {Rev. Sci.
		Instrum.}\ }\textbf {\bibinfo {volume} {68}},\ \bibinfo {pages} {3477}
(\bibinfo {year} {1997})}\BibitemShut {NoStop}%
\bibitem [{\citenamefont {K\"{o}hler}\ \emph {et~al.}(2011)\citenamefont
	{K\"{o}hler}, \citenamefont {Wollenhaupt}, \citenamefont {Bayer},
	\citenamefont {Sarpe},\ and\ \citenamefont
	{Baumert}}]{Koehler:2011:OE:11638}%
\BibitemOpen
\bibfield  {author} {\bibinfo {author} {\bibfnamefont {J.}~\bibnamefont
		{K\"{o}hler}}, \bibinfo {author} {\bibfnamefont {M.}~\bibnamefont
		{Wollenhaupt}}, \bibinfo {author} {\bibfnamefont {T.}~\bibnamefont {Bayer}},
	\bibinfo {author} {\bibfnamefont {C.}~\bibnamefont {Sarpe}},\ and\ \bibinfo
	{author} {\bibfnamefont {T.}~\bibnamefont {Baumert}},\ }\bibfield  {title}
{\bibinfo {title} {Zeptosecond precision pulse shaping},\ }\href@noop {}
{\bibfield  {journal} {\bibinfo  {journal} {Opt. Express}\ }\textbf {\bibinfo
		{volume} {19}},\ \bibinfo {pages} {11638} (\bibinfo {year}
	{2011})}\BibitemShut {NoStop}%
\bibitem [{\citenamefont {Wituschek}\ \emph {et~al.}(2016)\citenamefont
	{Wituschek}, \citenamefont {von Vangerow}, \citenamefont {Grzesiak},
	\citenamefont {Stienkemeier},\ and\ \citenamefont
	{Mudrich}}]{Wituschek:2016:RSI:083105}%
\BibitemOpen
\bibfield  {author} {\bibinfo {author} {\bibfnamefont {A.}~\bibnamefont
		{Wituschek}}, \bibinfo {author} {\bibfnamefont {J.}~\bibnamefont {von
			Vangerow}}, \bibinfo {author} {\bibfnamefont {J.}~\bibnamefont {Grzesiak}},
	\bibinfo {author} {\bibfnamefont {F.}~\bibnamefont {Stienkemeier}},\ and\
	\bibinfo {author} {\bibfnamefont {M.}~\bibnamefont {Mudrich}},\ }\bibfield
{title} {\bibinfo {title} {A simple photoionization scheme for characterizing
		electron and ion spectrometers},\ }\href@noop {} {\bibfield  {journal}
	{\bibinfo  {journal} {Rev. Sci. Instrum.}\ }\textbf {\bibinfo {volume}
		{87}},\ \bibinfo {pages} {083105} (\bibinfo {year} {2016})}\BibitemShut
{NoStop}%
\bibitem [{\citenamefont {Baumert}\ \emph {et~al.}(1997)\citenamefont
	{Baumert}, \citenamefont {Brixner}, \citenamefont {Seyfried}, \citenamefont
	{Strehle},\ and\ \citenamefont {Gerber}}]{Baumert:1997:APB:779}%
\BibitemOpen
\bibfield  {author} {\bibinfo {author} {\bibfnamefont {T.}~\bibnamefont
		{Baumert}}, \bibinfo {author} {\bibfnamefont {T.}~\bibnamefont {Brixner}},
	\bibinfo {author} {\bibfnamefont {V.}~\bibnamefont {Seyfried}}, \bibinfo
	{author} {\bibfnamefont {M.}~\bibnamefont {Strehle}},\ and\ \bibinfo {author}
	{\bibfnamefont {G.}~\bibnamefont {Gerber}},\ }\bibfield  {title} {\bibinfo
	{title} {Femtosecond pulse shaping by an evolutionary algorithm with
		feedback},\ }\href@noop {} {\bibfield  {journal} {\bibinfo  {journal} {Appl.
			Phys. B}\ }\textbf {\bibinfo {volume} {65}},\ \bibinfo {pages} {779}
	(\bibinfo {year} {1997})}\BibitemShut {NoStop}%
\bibitem [{\citenamefont {Yelin}\ \emph {et~al.}(1997)\citenamefont {Yelin},
	\citenamefont {Meshulach},\ and\ \citenamefont
	{Silberberg}}]{Yelin:1997:OL:1793}%
\BibitemOpen
\bibfield  {author} {\bibinfo {author} {\bibfnamefont {D.}~\bibnamefont
		{Yelin}}, \bibinfo {author} {\bibfnamefont {D.}~\bibnamefont {Meshulach}},\
	and\ \bibinfo {author} {\bibfnamefont {Y.}~\bibnamefont {Silberberg}},\
}\bibfield  {title} {\bibinfo {title} {Adaptive femtosecond pulse
	compression},\ }\href@noop {} {\bibfield  {journal} {\bibinfo  {journal}
	{Opt. Lett.}\ }\textbf {\bibinfo {volume} {22}},\ \bibinfo {pages} {1793}
(\bibinfo {year} {1997})}\BibitemShut {NoStop}%
\bibitem [{\citenamefont {Weiner}(2000)}]{Weiner:2000:RSI:1929}%
\BibitemOpen
\bibfield  {author} {\bibinfo {author} {\bibfnamefont {A.~M.}\ \bibnamefont
		{Weiner}},\ }\bibfield  {title} {\bibinfo {title} {Femtosecond pulse shaping
		using spatial light modulators},\ }\href@noop {} {\bibfield  {journal}
	{\bibinfo  {journal} {Rev. Sci. Instrum.}\ }\textbf {\bibinfo {volume}
		{71}},\ \bibinfo {pages} {1929} (\bibinfo {year} {2000})}\BibitemShut
{NoStop}%
\bibitem [{\citenamefont {Meshulach}\ and\ \citenamefont
	{Silberberg}(1998)}]{Meshulach:1998:Nature:239}%
\BibitemOpen
\bibfield  {author} {\bibinfo {author} {\bibfnamefont {D.}~\bibnamefont
		{Meshulach}}\ and\ \bibinfo {author} {\bibfnamefont {Y.}~\bibnamefont
		{Silberberg}},\ }\bibfield  {title} {\bibinfo {title} {Coherent quantum
		control of two-photon transitions by a femtosecond laser pulse},\ }\href@noop
{} {\bibfield  {journal} {\bibinfo  {journal} {Nature}\ }\textbf {\bibinfo
		{volume} {396}},\ \bibinfo {pages} {239} (\bibinfo {year}
	{1998})}\BibitemShut {NoStop}%
\bibitem [{\citenamefont {Wollenhaupt}\ \emph {et~al.}(2006)\citenamefont
	{Wollenhaupt}, \citenamefont {Pr\"{a}kelt}, \citenamefont {Sarpe-Tudoran},
	\citenamefont {Liese}, \citenamefont {Bayer},\ and\ \citenamefont
	{Baumert}}]{Wollenhaupt:2006:PRA:063409}%
\BibitemOpen
\bibfield  {author} {\bibinfo {author} {\bibfnamefont {M.}~\bibnamefont
		{Wollenhaupt}}, \bibinfo {author} {\bibfnamefont {A.}~\bibnamefont
		{Pr\"{a}kelt}}, \bibinfo {author} {\bibfnamefont {C.}~\bibnamefont
		{Sarpe-Tudoran}}, \bibinfo {author} {\bibfnamefont {D.}~\bibnamefont
		{Liese}}, \bibinfo {author} {\bibfnamefont {T.}~\bibnamefont {Bayer}},\ and\
	\bibinfo {author} {\bibfnamefont {T.}~\bibnamefont {Baumert}},\ }\bibfield
{title} {\bibinfo {title} {Femtosecond strong-field quantum control with
		sinusoidally phase-modulated pulses},\ }\href@noop {} {\bibfield  {journal}
	{\bibinfo  {journal} {Phys. Rev. A}\ }\textbf {\bibinfo {volume} {73}},\
	\bibinfo {pages} {063409} (\bibinfo {year} {2006})}\BibitemShut {NoStop}%
\bibitem [{\citenamefont {Pengel}\ \emph
	{et~al.}(2017{\natexlab{b}})\citenamefont {Pengel}, \citenamefont
	{Kerbstadt}, \citenamefont {Englert}, \citenamefont {Bayer},\ and\
	\citenamefont {Wollenhaupt}}]{Pengel:2017:PRA:043426}%
\BibitemOpen
\bibfield  {author} {\bibinfo {author} {\bibfnamefont {D.}~\bibnamefont
		{Pengel}}, \bibinfo {author} {\bibfnamefont {S.}~\bibnamefont {Kerbstadt}},
	\bibinfo {author} {\bibfnamefont {L.}~\bibnamefont {Englert}}, \bibinfo
	{author} {\bibfnamefont {T.}~\bibnamefont {Bayer}},\ and\ \bibinfo {author}
	{\bibfnamefont {M.}~\bibnamefont {Wollenhaupt}},\ }\bibfield  {title}
{\bibinfo {title} {Control of three-dimensional electron vortices from
		femtosecond multiphoton ionization},\ }\href@noop {} {\bibfield  {journal}
	{\bibinfo  {journal} {Phys. Rev. A}\ }\textbf {\bibinfo {volume} {96}},\
	\bibinfo {pages} {043426} (\bibinfo {year} {2017}{\natexlab{b}})}\BibitemShut
{NoStop}%
\bibitem [{\citenamefont {Kerbstadt}\ \emph
	{et~al.}(2019{\natexlab{b}})\citenamefont {Kerbstadt}, \citenamefont
	{Eickhoff}, \citenamefont {Bayer},\ and\ \citenamefont
	{Wollenhaupt}}]{Kerbstadt:2019:APX:1672583}%
\BibitemOpen
\bibfield  {author} {\bibinfo {author} {\bibfnamefont {S.}~\bibnamefont
		{Kerbstadt}}, \bibinfo {author} {\bibfnamefont {K.}~\bibnamefont {Eickhoff}},
	\bibinfo {author} {\bibfnamefont {T.}~\bibnamefont {Bayer}},\ and\ \bibinfo
	{author} {\bibfnamefont {M.}~\bibnamefont {Wollenhaupt}},\ }\bibfield
{title} {\bibinfo {title} {Control of free electron wave packets by
		polarization-tailored ultrashort bichromatic laser fields},\ }\href@noop {}
{\bibfield  {journal} {\bibinfo  {journal} {Adv. Phys.: X}\ }\textbf
	{\bibinfo {volume} {4}},\ \bibinfo {pages} {1672583} (\bibinfo {year}
	{2019}{\natexlab{b}})}\BibitemShut {NoStop}%
\bibitem [{\citenamefont {Eickhoff}\ \emph
	{et~al.}(2021{\natexlab{c}})\citenamefont {Eickhoff}, \citenamefont {Feld},
	\citenamefont {K\"{o}hnke}, \citenamefont {Englert}, \citenamefont {Bayer},\
	and\ \citenamefont {Wollenhaupt}}]{Eickhoff:2021:JPBAMOP:164002}%
\BibitemOpen
\bibfield  {author} {\bibinfo {author} {\bibfnamefont {K.}~\bibnamefont
		{Eickhoff}}, \bibinfo {author} {\bibfnamefont {L.}~\bibnamefont {Feld}},
	\bibinfo {author} {\bibfnamefont {D.}~\bibnamefont {K\"{o}hnke}}, \bibinfo
	{author} {\bibfnamefont {L.}~\bibnamefont {Englert}}, \bibinfo {author}
	{\bibfnamefont {T.}~\bibnamefont {Bayer}},\ and\ \bibinfo {author}
	{\bibfnamefont {M.}~\bibnamefont {Wollenhaupt}},\ }\bibfield  {title}
{\bibinfo {title} {Coherent control mechanisms in bichromatic multiphoton
		ionization},\ }\href@noop {} {\bibfield  {journal} {\bibinfo  {journal}
		{Journal of Physics B: Atomic, Molecular and Optical Physics}\ }\textbf
	{\bibinfo {volume} {54}},\ \bibinfo {pages} {164002} (\bibinfo {year}
	{2021}{\natexlab{c}})}\BibitemShut {NoStop}%
\bibitem [{\citenamefont {Feit}\ and\ \citenamefont
	{Fleck~Jr}(1974)}]{Feit:1974:APL:169}%
\BibitemOpen
\bibfield  {author} {\bibinfo {author} {\bibfnamefont {M.~D.}\ \bibnamefont
		{Feit}}\ and\ \bibinfo {author} {\bibfnamefont {J.~A.}\ \bibnamefont
		{Fleck~Jr}},\ }\bibfield  {title} {\bibinfo {title} {Effect of refraction on
		spot-size dependence of laser-induced breakdown},\ }\href@noop {} {\bibfield
	{journal} {\bibinfo  {journal} {Appl. Phys. Lett.}\ }\textbf {\bibinfo
		{volume} {24}},\ \bibinfo {pages} {169} (\bibinfo {year} {1974})}\BibitemShut
{NoStop}%
\bibitem [{\citenamefont {Cohen-Tannoudji}\ \emph {et~al.}(1977)\citenamefont
	{Cohen-Tannoudji}, \citenamefont {Diu},\ and\ \citenamefont
	{Laloe}}]{Cohen-Tannoudji:1977:3}%
\BibitemOpen
\bibfield  {author} {\bibinfo {author} {\bibfnamefont {C.}~\bibnamefont
		{Cohen-Tannoudji}}, \bibinfo {author} {\bibfnamefont {B.}~\bibnamefont
		{Diu}},\ and\ \bibinfo {author} {\bibfnamefont {F.}~\bibnamefont {Laloe}},\
}\href@noop {} {\emph {\bibinfo {title} {Quantum Mechanics Vol. 1}}}\
(\bibinfo  {publisher} {Wiley},\ \bibinfo {year} {1977})\ pp.\ \bibinfo
{pages} {3--889}\BibitemShut {NoStop}%
\bibitem [{\citenamefont {Wollenhaupt}\ \emph {et~al.}(2010)\citenamefont
	{Wollenhaupt}, \citenamefont {Bayer}, \citenamefont {Vitanov},\ and\
	\citenamefont {Baumert}}]{Wollenhaupt:2010:PRA:053422}%
\BibitemOpen
\bibfield  {author} {\bibinfo {author} {\bibfnamefont {M.}~\bibnamefont
		{Wollenhaupt}}, \bibinfo {author} {\bibfnamefont {T.}~\bibnamefont {Bayer}},
	\bibinfo {author} {\bibfnamefont {N.~V.}\ \bibnamefont {Vitanov}},\ and\
	\bibinfo {author} {\bibfnamefont {T.}~\bibnamefont {Baumert}},\ }\bibfield
{title} {\bibinfo {title} {Three-state selective population of dressed states
		via generalized spectral phase-step modulation},\ }\href@noop {} {\bibfield
	{journal} {\bibinfo  {journal} {Phys. Rev. A}\ }\textbf {\bibinfo {volume}
		{81}},\ \bibinfo {pages} {053422} (\bibinfo {year} {2010})}\BibitemShut
{NoStop}%
\bibitem [{\citenamefont {Meshulach}\ and\ \citenamefont
	{Silberberg}(1999)}]{Meshulach:1999:PRA:1287}%
\BibitemOpen
\bibfield  {author} {\bibinfo {author} {\bibfnamefont {D.}~\bibnamefont
		{Meshulach}}\ and\ \bibinfo {author} {\bibfnamefont {Y.}~\bibnamefont
		{Silberberg}},\ }\bibfield  {title} {\bibinfo {title} {Coherent quantum
		control of multiphoton transitions by shaped ultrashort optical pulses},\
}\href@noop {} {\bibfield  {journal} {\bibinfo  {journal} {Phys. Rev. A}\
}\textbf {\bibinfo {volume} {60}},\ \bibinfo {pages} {1287} (\bibinfo {year}
{1999})}\BibitemShut {NoStop}%
\bibitem [{\citenamefont {Cohen-Tannoudji}\ \emph {et~al.}(2019)\citenamefont
	{Cohen-Tannoudji}, \citenamefont {Diu},\ and\ \citenamefont
	{Lalo\"{e}}}]{Cohen-Tannoudji:2019}%
\BibitemOpen
\bibfield  {author} {\bibinfo {author} {\bibfnamefont {C.}~\bibnamefont
		{Cohen-Tannoudji}}, \bibinfo {author} {\bibfnamefont {B.}~\bibnamefont
		{Diu}},\ and\ \bibinfo {author} {\bibfnamefont {F.}~\bibnamefont
		{Lalo\"{e}}},\ }\href@noop {} {\emph {\bibinfo {title} {Quantum Mechanics,
			Volume 3: Fermions, Bosons, Photons, Correlations, and Entanglement}}}\
(\bibinfo  {publisher} {John Wiley \& Sons},\ \bibinfo {year}
{2019})\BibitemShut {NoStop}%
\bibitem [{\citenamefont {Dudovich}\ \emph {et~al.}(2002)\citenamefont
	{Dudovich}, \citenamefont {Oron},\ and\ \citenamefont
	{Silberberg}}]{Dudovich:2002:PRL:123004}%
\BibitemOpen
\bibfield  {author} {\bibinfo {author} {\bibfnamefont {N.}~\bibnamefont
		{Dudovich}}, \bibinfo {author} {\bibfnamefont {D.}~\bibnamefont {Oron}},\
	and\ \bibinfo {author} {\bibfnamefont {Y.}~\bibnamefont {Silberberg}},\
}\bibfield  {title} {\bibinfo {title} {Coherent transient enhancement of
	optically induced resonant transitions},\ }\href@noop {} {\bibfield
{journal} {\bibinfo  {journal} {Phys. Rev. Lett.}\ }\textbf {\bibinfo
	{volume} {88}},\ \bibinfo {pages} {123004 } (\bibinfo {year}
{2002})}\BibitemShut {NoStop}%
\bibitem [{\citenamefont {Osgood}(2019)}]{Osgood:2019}%
\BibitemOpen
\bibfield  {author} {\bibinfo {author} {\bibfnamefont {B.~G.}\ \bibnamefont
		{Osgood}},\ }\href@noop {} {\emph {\bibinfo {title} {Lectures on the Fourier
			Transform and its Applications}}},\ Vol.~\bibinfo {volume} {33}\ (\bibinfo
{publisher} {American Mathematical Soc.},\ \bibinfo {address} {Providence,
	Rhode Island},\ \bibinfo {year} {2019})\BibitemShut {NoStop}%
\bibitem [{\citenamefont {Bracewell}(2000)}]{Bracewell:2000}%
\BibitemOpen
\bibfield  {author} {\bibinfo {author} {\bibfnamefont {R.}~\bibnamefont
		{Bracewell}},\ }\href@noop {} {\emph {\bibinfo {title} {The Fourier Transform
			and Its Applications}}},\ Vol.\ \bibinfo {volume} {3rd}\ (\bibinfo
{publisher} {McGraw-Hill Higher Education},\ \bibinfo {address} {Singapore},\
\bibinfo {year} {2000})\BibitemShut {NoStop}%
\bibitem [{\citenamefont {Goodman}(1996)}]{Goodman:1996:xi}%
\BibitemOpen
\bibfield  {author} {\bibinfo {author} {\bibfnamefont {J.~W.}\ \bibnamefont
		{Goodman}},\ }\href@noop {} {\emph {\bibinfo {title} {Introduction to Fourier
			Optics}}},\ Vol.~\bibinfo {volume} {2}\ (\bibinfo  {publisher} {THE
	McGRAW-HILL COMPANIES, INC.},\ \bibinfo {year} {1996})\ pp.\ \bibinfo {pages}
{xi--441}\BibitemShut {NoStop}%
\bibitem [{\citenamefont {Dudovich}\ \emph {et~al.}(2005)\citenamefont
	{Dudovich}, \citenamefont {Polack}, \citenamefont {Pe'er},\ and\
	\citenamefont {Silberberg}}]{Dudovich:2005:PRL:083002}%
\BibitemOpen
\bibfield  {author} {\bibinfo {author} {\bibfnamefont {N.}~\bibnamefont
		{Dudovich}}, \bibinfo {author} {\bibfnamefont {T.}~\bibnamefont {Polack}},
	\bibinfo {author} {\bibfnamefont {A.}~\bibnamefont {Pe'er}},\ and\ \bibinfo
	{author} {\bibfnamefont {Y.}~\bibnamefont {Silberberg}},\ }\bibfield  {title}
{\bibinfo {title} {Simple route to strong-field coherent control},\
}\href@noop {} {\bibfield  {journal} {\bibinfo  {journal} {Phys. Rev. Lett.}\
}\textbf {\bibinfo {volume} {94}},\ \bibinfo {pages} {083002 } (\bibinfo
{year} {2005})}\BibitemShut {NoStop}%
\end{thebibliography}

%

\end{document}